\documentclass[10pt,prl,aps,twocolumn,superscriptaddress,preprintnumbers]{revtex4-1}
\usepackage{graphicx, subfigure}
\usepackage{amssymb}
\usepackage{amsmath}
\usepackage{bm}
\usepackage{tikz}

\newcommand{\bc}{\begin{center}}
\newcommand{\ec}{\end{center}}
\def\ba#1{\begin{array}{#1}\displaystyle}
\newcommand{\ea}{\end{array}}

\newcommand{\beq}{\begin{equation}}
\newcommand{\eeq}{\end{equation}}
\newcommand{\beqa}{\begin{eqnarray}}
\newcommand{\eeqa}{\end{eqnarray}}

\newcommand{\bi}{\begin{itemize}}
\newcommand{\ei}{\end{itemize}}

\newcommand{\Tr}{{\rm Tr}}

\newcommand{\Or}{{\cal O}}

\begin{document}

\title{Entanglement Oscillations near a Quantum Critical Point}

\author{Olalla A. Castro-Alvaredo}
\affiliation
{Department of Mathematics, City, University of London, 10 Northampton Square, EC1V 0HB London, UK}
\author{M\'at\'e Lencs\'es}
\affiliation
{BME Department of Theoretical Physics,  H-1111 Budapest, Budafoki \'ut 8.}
\affiliation
{BME ``Momentum'' Statistical Field Theory Research Group, 
H-1111 Budapest, Budafoki \'ut 8.}
\author{Istv\'an M. Sz\'ecs\'enyi}
\affiliation
{Nordita, KTH Royal Institute of Technology and Stockholm University, Roslagstullsbacken 23, SE-106 91 Stockholm, Sweden}
\author{Jacopo Viti}
\affiliation
{International Institute of Physics \& ECT, UFRN, Campos Universit\'ario, Lagoa Nova 59078-970 Natal, Brazil}
\affiliation
{INFN, Sezione di Firenze, Via G. Sansone 1, 50019 Sesto Fiorentino, Firenze, Italy}
\date{\today}

\begin{abstract} 
We study the dynamics of entanglement in the scaling limit of the Ising spin chain in the presence of both a longitudinal and a transverse field.
We present analytical results for the quench of the longitudinal field in  critical transverse field which go beyond current lattice integrability techniques. 
We test these results against a numerical simulation on the corresponding lattice model finding extremely good agreement.
We show that the presence of bound states in the spectrum of the field theory leads to oscillations in the entanglement entropy and suppresses its linear growth on the time scales accessible to numerical simulations. For small quenches we determine exactly these oscillatory contributions and demonstrate that their presence follows from symmetry arguments. 
For the quench of the transverse field at zero longitudinal field we prove that the R\'enyi entropies are exactly proportional to the logarithm of the exponential of a time-dependent function, whose leading large-time behaviour is linear, hence entanglement grows linearly. 
We conclude that, in the scaling limit, linear growth and oscillations in the entanglement entropies can not be simply seen as consequences of integrability and its breaking respectively.

\end{abstract}

\preprint{NORDITA 2020-006}

\maketitle

{\em Introduction.---} Over the past two decades, one-dimensional many-body quantum systems far from equilibrium have become ubiquitous laboratories to scrutinize fundamental aspects of statistical mechanics.
 Out-of-equilibrium protocols featuring unitary dynamics, such as quantum quenches, have been commonly employed to test relaxation and thermalization  hypothesis~\cite{ETH2, ETH3}~ in experimentally realizable setups~\cite{kinoshita, Schmidt_07, Gring1318, Langen207}.  
A powerful theoretical device that tests whether a physical system can eventually approach equilibrium is represented by its entanglement dynamics~\cite{EEquench}. In 1+1 dimensions, the linear-in-time increase of the entanglement entropies is a signature that local observables relax  exponentially fast~\cite{quench, CEF, FE} and thermalize~\cite{Essler2016, Mori_2018}. This characteristic growth has been further  conjectured to be a generic feature of integrable models~\cite{AC}, where it has been analyzed within a quasi-particle picture,  inspired by conformal field theory~\cite{EEquench} and free fermion calculations~\cite{FC}. In such a framework, entangled quasi-particle pairs propagate freely in space-time and  generate linearly growing  entropies. Minimal models, with random unitary dynamics, that show analogous entanglement growth, have been also studied~\cite{Nahum_2016, PhysRevX.8.041019}  in connection with quantum chaos~\cite{PhysRevX.8.021062, PhysRevX.9.021033} and non-integrable systems.

Nonetheless, there exist  a vast class of one-dimensional systems that escape this paradigm and fail to relax at large times after the quench. 
They have been observed both in earlier studies~\cite{PhysRevLett.106.050405} and in actual recent experiments~\cite{scar_Exp}, see also~\cite{Abanin}. Within a qualitative quasi-particle picture~\cite{Gabor1},  absence of thermalization has been associated with integrability breaking interactions and confinement.
Numerical studies in the Ising spin chain~\cite{Gabor1} and its scaling limit~\cite{RAKOVSZKY2016805, 10.21468/SciPostPhys.5.3.027, Konik1, Konik2} indicated that in the presence of a longitudinal field  entanglement growth is strongly suppressed while local observables  feature persistent oscillations whose frequencies coincide with the  meson masses.
A similar lack of relaxation  has been also found later in a variety of physical models spanning from  gauge theories~\cite{C1, C3, Cubero2, Magnifico} and fractons~\cite{C2} to Heisenberg magnets~\cite{TC} and systems with long-range interactions~\cite{C4}.
However, despite a large number of numerical investigations, most of the understanding of whether  and how local observables will  equilibrate after a quench remains at a phenomenological level. This is mainly  because no lattice integrability technique~\cite{korepinbook} is available to systematically analyze these strongly interacting systems. 

In this Letter, we put forward a unified  picture to address perturbatively questions about  entanglement dynamics and relaxation in gapped  1+1  dimensional systems close to a Quantum Critical Point (QCP).
In particular, for the first time, we provide an analytical grasp on how entanglement growth can be so dramatically different depending on the non-equilibrium protocol considered. The formalism combines the perturbative approach of~\cite{PQ1,PQ2} with the mapping in the scaling limit between powers of the reduced density matrix and correlation functions of a local field, called the branch point twist field~\cite{Calabrese:2004eu, entropy}. For massive systems, this mapping has been successfully employed at equilibrium~\cite{review} and recently also in a time-dependent context~\cite{ourIsing}.  Crucially, its conclusions do not rely on any \textit{a priori}  assumption about the space-time evolution of the quasi-particles.

Focussing on the illustrative example of a quench of the longitudinal field in the ferromagnetic Ising spin chain, we will provide   examples of how  bound state formation and symmetries of the twist field are responsible for slow relaxation of local observables and  oscillations in the entanglement entropies.  Although derived through field theory techniques our results are tested numerically in the lattice model in the scaling limit and very good agreement is found.

{\em Model.---}
Consider the  ferromagnetic  Ising spin chain defined by the  Hamiltonian
\begin{equation}
\label{Ising_ham}
 H_{\text{lattice}}=-\sum_{n\in\mathbb Z}\left[\sigma^{x}_{n}\sigma^{x}_{n+1}+h_{z}\sigma^{z}_{n}+h_{x}\sigma^{x}_{n}\right]\,.
\end{equation}
The  Ising  chain is a prototype of a quantum phase transition with spontaneous breaking of $\mathbb Z_2$ symmetry and  is critical for $h_{z}=1$ and $h_{x}=0$. At criticality, the low energy excitations are massless free Majorana fermions described by a conformal field theory with central charge $c=1/2$~\cite{BPZ}. Within the renormalization group framework, near the QCP,  expectation values of local operators in the spin chain can be calculated from the relativistic Quantum Field Theory (QFT) action
\begin{equation}
 \label{action}
 \mathcal{A}_0=\mathcal{A}^{\rm{CFT}}-\lambda_1\int \mathrm{d}x~\mathrm{d}t ~\varepsilon(x,t)-\lambda_2\int \mathrm{d}x~\mathrm{d}t~\sigma(x,t)\,,
\end{equation}
which is the celebrated Ising field theory (IFT)~\cite{Zamolodchikov:1989fp, Delfino:2003yr, qft}. In Eq.~\eqref{action}, the conformal invariant action $\mathcal{A}^{\rm{CFT}}$ is perturbed by the $\mathbb Z_2$ even field $\varepsilon$  (energy), which is the continuum version of the lattice operator $\sigma^{z}_{n}$, and the $\mathbb Z_2$ odd field $\sigma$ (spin), which is instead  the continuum version of the order parameter $\sigma_{n}^{x}$. The coupling constant $\lambda_1$ is proportional to the deviation of the transverse field from its critical value $h_{z}-1$, while $\lambda_2$ is proportional to the longitudinal field $h_{x}$. At the QCP, the scaling dimension of $\sigma$ is $\Delta_{\sigma}=1/8$ and that of $\varepsilon$ is $\Delta_{\varepsilon}=1$. 

 Let $|\Omega\rangle$ be the ground state of the Hamiltonian $H$ of the field theory ~\eqref{action}. Following a widely studied non-equilibrium protocol, dubbed  \textit{quantum quench},  at time $t=0$  one of the two coupling constants $\lambda_{i}$ ($i=1,2$) is modified according to $\lambda_{i}\rightarrow\lambda_i+\delta_\lambda$.  The evolution of the pre-quench ground state $|\Omega\rangle$  is governed by the perturbed Hamiltonian
\begin{equation}
\label{post}
G(t):=H+\theta(t)\delta_\lambda\int \mathrm{d}x~\Psi(x)\,,
\end{equation}
$\Psi$ being either the spin or the energy field  and  $\theta(t)$, the Heaviside step function.
This dynamical problem is  analytically not solvable in general~\cite{PQ1}. To provide theoretical predictions for local observables and entanglement entropies following a quench, one thus sets up a perturbative expansion in the relative quench parameter $\frac{\delta _\lambda}{\lambda_i}\ll 1$.

{\em Perturbation Theory.---} We revisit and extend to include entanglement calculations, the perturbative approach to the quench problem~\cite{PQ1}. In a relativistic   scattering theory, it is  possible to consider a basis of in and out states, denoted by $|\alpha\rangle^{\text{in/out}}$, which are multi-particle eigenstates of the  Hamiltonian $H=G(-\infty)$. In particular, a single-particle eigenstate of $H$ has energy $e(p)=\sqrt{m_0^2+p^2}$, where $m_0$ is its pre-quench mass and $p$ the momentum. Similar eigenbases are constructed for the post-quench Hamiltonian  $H_{\text{post}}:= H+\delta_\lambda\int \mathrm{d}x~\Psi(x)=G(\infty)$. In this case, the energy of a  single-particle state will be denoted by $\tilde{e}(p)=\sqrt{m^2+p^2}$, being $m$ its post-quench mass.

The initial state $|\Omega\rangle$ can be formally  expanded into the basis of  the out-states of $H_{\text{post}}$  as:
$|\Omega\rangle=\sum_{\alpha}c_{\alpha}|\alpha\rangle_{\text{post}}^{\text{out}}$. Assuming for simplicity a unique family of particles, $|\alpha\rangle_{\text{post}}^{\text{out}}$  is then the  multiparticle out-state $|p_{1},\dots,p_n\rangle_{\text{post}}^{\text{out}}$ while the symbol  $\sum_{\alpha}$ is a shorthand notation for~the Lorentz invariant integration measure in 1+1 dimensions~\cite{foot}. The overlap coefficients $c_{\alpha}$ are the elements of the scattering matrix for the quench problem in Eq.~\eqref{post}.
At first-order in perturbation theory, one has~\cite{PIM, Overlap1} 
\begin{equation}
\label{expansion}
 |\Omega\rangle=|\Omega\rangle_{\text{post}}+2\pi\delta_\lambda\sum_{\alpha\not=\Omega}
 \frac{\delta(P^{\alpha})}{E^{\alpha}}~(F_{\alpha}^{\Psi})^{*}|\alpha\rangle_{\text{post}}^{\text{out}}+\Or(\delta_{\lambda}^2)\,.
\end{equation}
In Eq.~\eqref{expansion}, $E^{\alpha}$ and $P^{\alpha}$ are the pre-quench energy and momentum of the state $|\alpha\rangle^{\text{out}}$; $\delta(x)$ is the Dirac delta and the function $F^{\Psi}_{\alpha}$ is the  form factor: $F^{\Psi}_{\alpha}:=\langle\Omega|\Psi(0,0)|\alpha\rangle^{\text{in}}$, calculated in the pre-quench theory. 
From the expansion in Eq.~\eqref{expansion}, it is straightforward to derive the post-quench evolution of a local operator $\Phi$
 \begin{eqnarray}
 \label{int_pert}
  &&\langle\Omega|\Phi(0,t)|\Omega\rangle={}_{\text{post}}\langle\Omega|\Phi(0,0)|\Omega\rangle_{\text{post}}\\
  && \nonumber
  +4\pi\delta_{\lambda}\sum_{\alpha\not=\Omega}~\frac{\delta(P^{\alpha})}{E^{\alpha}}\text{Re}\left[e^{-itE_{\text{post}}^{\alpha}}(F_{\alpha}^{\Psi})^{*}~\langle\Omega|\Phi(0,0)|\alpha\rangle_{\text{post}}^{\text{out}}\right]\\\nonumber
  &&+\Or(\delta_{\lambda}^2)\,, \qquad\qquad\qquad\qquad\qquad\qquad\qquad\qquad\qquad
 \end{eqnarray}
with now $E^{\alpha}_{\text{post}}$ the energy of the  state $|\alpha\rangle_{\text{post}}^{\text{out}}$. At $\Or(\delta_{\lambda})$, one can replace $|\alpha\rangle_{\text{post}}^{\text{out}}$ by $|\alpha\rangle^{\text{out}}$ inside the sum in Eq.~\eqref{int_pert} and, by using known properties of the form factors, is also possible to relax the ordering prescription on the momenta of the out-states~\cite{smirnovbook}. We will denote then by $\sum_{\alpha}'$, a Lorentz invariant integration over the pre-quench multiparticle states with unrestricted momenta~\cite{Foot2}. The leading order correction to the one-point function of a local operator after a quench is therefore~\cite{PQ1, PQ2}
\begin{eqnarray}
\label{local}
&&\langle\Omega|\Phi(0,t)|\Omega\rangle={}_{\text{post}}\langle\Omega|\Phi(0,0)|\Omega\rangle_{\text{post}}\\
&&\nonumber+4\pi\delta_{\lambda}\sum_{\alpha\not=\Omega}'~\frac{\delta(P^{\alpha})}{E^{\alpha}}\text{Re}\left[e^{-itE^{\alpha}_{\text{post}}}(F_{\alpha}^{\Psi})^{*} F^{\Phi}_{\alpha}\right]+\Or(\delta_{\lambda}^2)\,.
\end{eqnarray}

Consider now  a semi-infinite  spatial bipartition of the Hilbert space of the QFT associated to the quench problem~\eqref{post}. In particular, let $\mathcal{L}$ be the semi-infinite negative real line and $\mathcal{R}$ the semi-infinite positive real line and denote by $\rho_{\mathcal{R}}(t):=\text{Tr}_{\mathcal{L}}[e^{-iH_{\text{post}}t}|\Omega\rangle\langle\Omega|e^{iH_{\text{post}}t}]$, the reduced density matrix after the quench obtained tracing over the left degrees of freedom. In QFT, half-space R\'enyi entropies after a quench $S_n(t):=\frac{1}{1-n}\log[\Tr_{\mathcal{R}} \rho_{\mathcal{R}}^n(t)]$ are related to the one-point function  of the  twist field $\mathcal{T}_n$~\cite{Calabrese:2004eu,entropy} by
 \begin{equation}
 \label{twist}
 S_n(t)=\frac{1}{1-n}\log\left[\epsilon^{ \Delta_{\mathcal{T}_n}}\langle \Omega|\mathcal{T}_n(0,t)|\Omega\rangle\right]\,.
 \end{equation}
 In Eq.~\eqref{twist}, $\epsilon$ is a short distance cut-off and $\Delta_{\mathcal{T}_n}=\frac{c}{12}(n-n^{-1})$ is the scaling dimension of the twist field at the QCP \cite{orbifold,Bouwknegt,Borisov}. The von Neumann entropy $S(t)$ is defined through the limit $S(t):=\lim_{n\rightarrow 1}S_n(t)$.
 
 In writing Eq.~\eqref{twist}, a new  difficulty arises: the expectation value of the twist field has to be calculated in an $n$-fold replicated QFT and  the time evolution of the twist field after the quench is governed by the replicated Hamiltonian $\sum_{r=1}^{n}H_{\text{post}}^{(r)}=\sum_{r=1}^{n}(H^{(r)}+\delta_{\lambda}\int dx\Psi^{(r)})$, $r$ being the replica index. 
 However, when calculating the  overlaps $_{\text{post}}^{\text{out}}\langle\alpha|\Omega\rangle$  at first-order in $\delta_{\lambda}$, the sum over the replica trivializes since the perturbing field $\Psi^{(r)}$ has non-vanishing matrix elements only between particles in the same copy. One therefore gets Eq.~\eqref{expansion} with a prefactor $n$ in front of the sum,  which only  involves states within one particular replica, for instance the first.
 By repeating now the derivation of Eq.~\eqref{local}, we conclude that the leading order expansion of the twist field one-point function after a quench is~
\begin{eqnarray}
\label{twist2}
&& \langle\Omega|\mathcal{T}_n(0,t)|\Omega\rangle={}_{\text{post}}\langle\Omega|\mathcal{T}_n(0,0)|\Omega\rangle_{\text{post}}\\
&& \!\!+4\pi n\delta_{\lambda}\!\!\!\!\!\!\sum_{\substack{\alpha\neq\Omega;\\ \alpha\in\text{1st rep.}}}'~\!\!\!\!\!\!\!\!\frac{\delta(P^{\alpha})}{E^{\alpha}}\text{Re}\left[e^{-itE_{\text{post}}^{\alpha}}(F_{\alpha}^{\Psi})^{*} F^{\mathcal{T}_n}_{\alpha}\right]+\Or(\delta_{\lambda}^2)\,,\nonumber
\end{eqnarray}
where, as indicated, the sum only contains states in the first replica. Similarly to the discussion around Eq.~\eqref{local}, $F^{\mathcal{T}_n}_{\alpha}$ in Eq.~\eqref{twist2} denotes the pre-quench twist-field matrix element $F_{\alpha}^{\mathcal{T}_n}=\langle\Omega|\mathcal{T}_{n}(0,0)|\alpha\rangle^{\text{in}}$.\\
\begin{figure}[h!]
\includegraphics[width=\columnwidth]{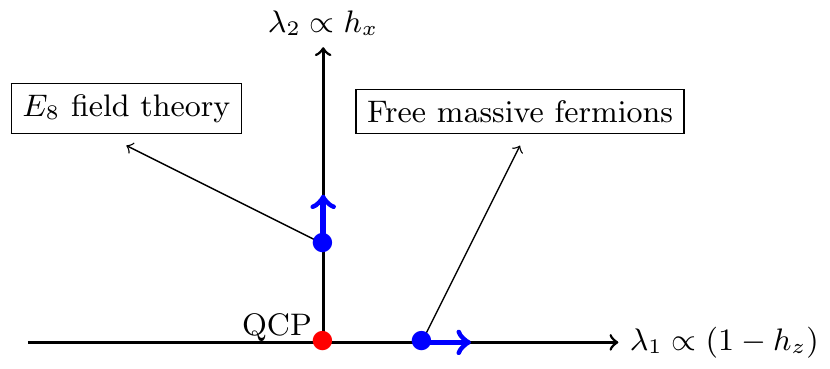}
\caption{Phase diagram of the IFT, described by the action ~\eqref{action}. We consider applications of the perturbation theory to a quench of the longitudinal field $h_x\propto\lambda_2$, while keeping the transverse field $(1-h_z)\propto\lambda_1$ at its critical value, i.e. $\lambda_1=0$. In the scaling limit, the pre-quench theory is integrable and corresponds to the $E_8$ field theory. For $\lambda_2=0$, the pre-quench theory can be mapped to non-interacting fermions with mass $\lambda_1$.}
\label{fig_quenches}
\end{figure}

{\em Longitudinal Field Quench.---} We examine  a quench along the vertical axis of the phase diagram of the IFT depicted in Fig.~\ref{fig_quenches}. This quench, cf. Eq.~\eqref{action}, involves a sudden change of the coupling $\lambda_2\rightarrow\lambda_2+\delta_{\lambda}$ while keeping $\lambda_1=0$. In the lattice model described by Eq.~\eqref{Ising_ham},  it modifies  the longitudinal field $h_x\rightarrow h_x+\delta_{h_x}$ at fixed transverse field $h_z=1$. In the presence of a longitudinal field, the Ising spin chain is strongly interacting and the perturbative approach is the only analytical device to  study   entanglement dynamics.

From a QFT perspective, at $\lambda_1=0$ and $\lambda_2\not=0$, both the pre- and post-quench theories are integrable. The spectrum contains of eight stable particles~\cite{Zamolodchikov:1989zs,Zamolodchikov:1989fp}, whose masses are in correspondence with the components of the Perron-Frobenius eigenvector of the Cartan matrix of the Lie algebra $E_8$. We will refer to such a field theory, in short, as the $E_8$ field theory, see Fig.~\ref{fig_quenches}. The masses of the eight particles have been partially measured experimentally~\cite{Coldea177} and numerically estimated in the scaling limit using matrix product states~\cite{PhysRevB.83.020407}. In the $E_8$ field theory both the spin operator and the twist field couple to the eight one-particle states. Eqs.~\eqref{local} and~\eqref{twist2}, predict in this case that at $\Or(\delta_{\lambda})$ the one-point function of the spin and the entanglement entropies must oscillate in time without relaxing. The first-order result for the order parameter~\cite{PQ2} is re-obtained in~\cite{SM}. For the time evolution of the entanglement entropies, perturbation theory, combined with Eq.~\eqref{twist}, gives at large times
\begin{multline}
\label{entropy_ev}
S_n(t)-S_n(0)\stackrel{t\gg 1}{=}\frac{\delta_\lambda}{\lambda_2}\left[\frac{2n\mathcal{C}_{\sigma}}{1-n}\sum_{a=1}^8\frac{\hat{F}^{\sigma}_{a}\hat{F}^{\mathcal{T}_n}_{a}}{r_{a}^2}\cos( m r_a t)\right.\\
+\left.\frac{1}{1-n}\frac{\Delta_{\mathcal{T}_n}}{2-\Delta_{\sigma}} \right]+\Or(\delta_{\lambda}^2)\,.
\end{multline}
where the coefficient~\cite{Fateev:1997yg} $\mathcal{C_{\sigma}}=-0.065841\dots$ and the (real)  normalized pre-quench one-particle form factors of the spin field~\cite{DM}, $\hat{F}^{\sigma}_{a}$, and the  twist field~\cite{E8toda}, $\hat{F}^{\mathcal{T}_n}_{a}$ are also summarized in~\cite{SM}.
 The universal ratios $r_a$ in Eq.~\eqref{entropy_ev} are the masses of the particles in the $E_8$ field theory normalized by the mass of the lightest particle, whose value \textit{after} the quench is $m$. It is finally possible~\cite{SM} to extrapolate the results for the R\'enyi entropies  to $n\rightarrow 1$, and predict  the long-time limit of the von Neumann entropy. There are  subleading corrections  in time to Eq.~\eqref{entropy_ev} of order $t^{-3/2}$ (but of leading order in $\delta_\lambda$) which  are  discussed in~\cite{SM}.

The field theoretical result in Eq.~\eqref{entropy_ev} can be tested against numerical simulations through matrix product states in the Ising spin chain near the QCP. For finding the initial state and for the time evolution we use the iTEBD algorithm~\cite{VidaliTEBD1,VidaliTEBD2} extrapolated to the scaling limit, details are given in~\cite{SM}. In a non-equilibrium protocol, the longitudinal field is quenched from $h_x$ to $h_x+\delta_{h_x}$ with $\delta_{h_x}/h_x=\delta_{\lambda}/\lambda_2=-0.04,0.05$. Due to the absence of visible linear growth of the entanglement entropies~\cite{SM}, the simulation can reach large enough time to carry out a Fourier analysis. The non-universal mass coupling relation is obtained by fitting the numerical data for the order parameter to the theoretical curve given~in~\cite{SM}, and we have $m\approx 5.42553 (h_x+\delta h_x)^{8/15}$, consistent with earlier estimates~\cite{Henkel_1989, PhysRevB.83.020407}.

\begin{figure}[t]
\includegraphics[width=\columnwidth]{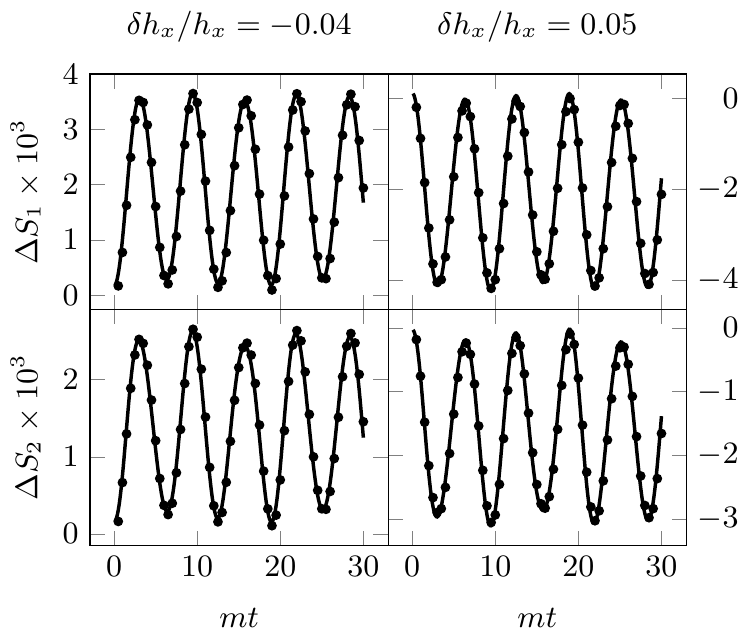}
\caption{The time evolution of the von Neumann entropy (top) and the second R\'enyi entropy (bottom) differences $\Delta S_n = S_n(t)-S_n(0)$ for quenches with $\delta h_x/h_x = -0.04$ (left) and $\delta h_x/h_x = 0.05$ (right). The dots are the extrapolated iTEBD data. Lines are the theoretical prediction from Eq.~\eqref{entropy_ev} ($n\rightarrow1$ limit for von Neumann), up to the first four particles in the sum, and incorporating the two particle contributions given in~\cite{SM}.}
\label{fig:ent_r}
\end{figure}

\begin{figure}[t]
\includegraphics[width=\columnwidth]{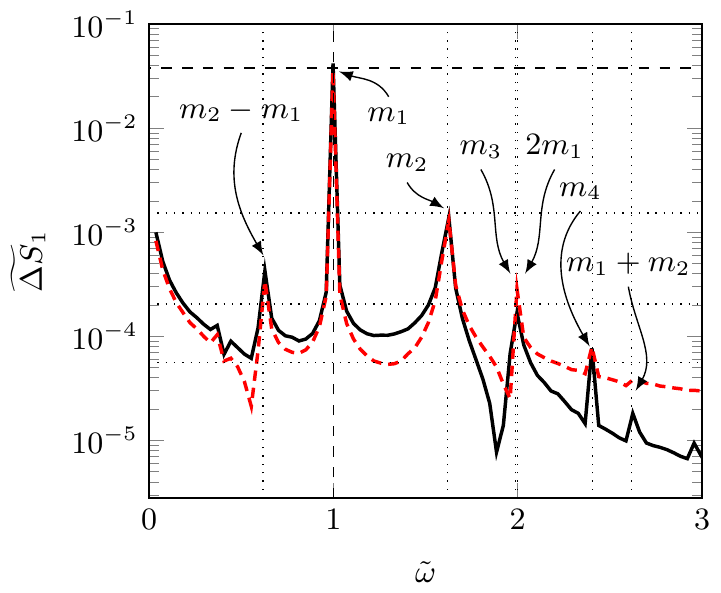}
\caption{Numerical Fourier transform of the variation of the von Neumann entropy (related to the variable $mt$) for quenches with $\delta h_x/h_x = 0.05$ (solid) and $\delta h_x/h_x = -0.04$ (dashed). Vertical lines indicate different frequencies.  The horizontal lines mark the peaks corresponding to the masses of the four lightest particles. From top to bottom they correspond to $m_1, m_2, m_3$ and $m_4$, respectively. The dashed horizontal line is set by hand, and the three dotted horizontal lines were calculated from the ratios of the one-particle form factors based on Eq.~\eqref{entropy_ev}.  \label{fig:ent_f}}
\end{figure}

According to Eq.~\eqref{entropy_ev}, in the scaling limit, with time measured in units of $m^{-1}$, the time evolution of entanglement 
entropies should follow a universal curve. The numerical results for real time evolution in the scaling region are  summarized in Fig.~\ref{fig:ent_r} for the von Neumann and the second R\'enyi entropy, showing excellent agreement with theoretical predictions  obtained from Eq.~\eqref{entropy_ev}. The curves for the entanglement entropies have been shifted vertically by an empirical value to account for higher order  corrections~\cite{ourIsing} to the twist field post-quench expectation value, cf. Eq.~(39) in~\cite{SM}. In Fig.~\ref{fig:ent_f} we also show the numerical Fourier spectrum of the von Neumann entropy calculated from extrapolated data up to $m t = 170$. The Fourier transform was carried out with respect to the rescaled time $mt$, therefore the main frequency is at $\tilde{\omega}=1$ for both quenches. The various peaks are related to the mass ratios of different particles summarized in~\cite{SM}. For infinite time the one particle peaks would be $\delta$-function peaks, but for finite time they have finite height. The height ratios are related to form factors of the longitudinal field and the twist fields through Eq.~\eqref{entropy_ev}. The horizontal line in Fig.~\ref{fig:ent_f} related to the lightest particle is set by hand, the ones related to $m_2,m_3$ and $m_4$ are calculated from the form factors given in~\cite{SM}.

{\em Transverse Field Quench.---} We consider now   a quench of the transverse field $h_z\rightarrow h_z+\delta_{h_z}$ for longitudinal field $h_x=0$.  In the IFT, see Fig.~\ref{fig_quenches},  this protocol displaces along the horizontal axis of the phase diagram: $\lambda_1\rightarrow\lambda_1+\delta_{\lambda}$, modifying the mass of  the Majorana fermion~\cite{CEF, SE}.
The ground state $|\Omega\rangle$ of the pre-quench theory can be expanded in the post-quench quasi-particle basis as
\begin{equation}
|\Omega\rangle=\text{exp}\left[\int_0^{\infty}\frac{\mathrm{d}p}{2\pi \tilde{e}(p)}\tilde{K}(p)a_{\text{post}}^{\dagger}(-p)a_{\text{post}}^{\dagger}(p)\right]|\Omega\rangle_{\text{post}}\,,
\label{eq:squeezed_vacuum_state}
\end{equation}
where the function $\tilde{K}(p)$ is given in~\cite{SE} and $a_{\text{post}}^{\dagger}(p)$  are post-quench fermionic creation operators. Due to the properties of the free fermionic form factors,  the expectation value of the twist field exponentiates
\begin{equation}
\label{exp}
\frac{\langle\Omega|\mathcal{T}_n(0,t)|\Omega\rangle}{{}_{\text{post}}\langle\Omega|\mathcal{T}_n(0,0)|\Omega\rangle_{\text{post}}}=\text{exp}\left[\sum_{k,l=0}^{\infty}D_{2k,2l}^{c}(t)\right]\,.
\end{equation}
For a proof of Eq.~\eqref{exp}, we  refer to the Supplementary Material~\cite{SM}.
The amplitudes $D^{c}_{2k,2l}(t)$ contribute at leading order $\left(\frac{\delta_{\lambda}}{\lambda_1}\right)^{k+l}$ in perturbation theory and can be systematically computed.
Differently from Eq.~\eqref{entropy_ev}, the oscillatory first-order term is $O(t^{-3/2})$ for large time, while in~\cite{ourIsing} it was  shown that $D^{c}_{2,2}(t)=\left(\frac{\delta_{\lambda}}{\lambda_1}\right)^2\bigl[-|A|t+\Or(1)\bigr]$.  In the absence of interactions,  exponentiation of 
second order contributions, leads then to linear growth of the R\'enyi entropies  as a by-product of  relaxation of the twist-field one-point function.

{\em Discussion.---}
Absence of relaxation of the order parameter and persistent  oscillations in the entanglement entropies  have been observed previously in several  numerical investigations of the Ising spin chain and its scaling limit~\cite{RAKOVSZKY2016805, Gabor1, Konik1, Konik2, 10.21468/SciPostPhys.5.3.027}.~ In this Letter,  we formulated a new first principle perturbative approach which quantitatively explains these phenomena.  Persistent oscillations in the one-point function of a local observable are only possible if it can create a single quasi-particle excitation of the post-quench Hamiltonian. This is a necessary condition that is never satisfied in absence of interactions, as also emphasized in~\cite{PQ1}. By mapping entanglement entropies into correlation functions of a local field, the twist field, we then provided an analogous criterion to understand when entanglement growth can slow down. 

An important question is whether the exponentiation of  higher orders in perturbation theory will  damp the oscillatory first-order result for local observables derived in Eqs.~\eqref{local} and (\ref{twist2}). 
 The time-scale for this to happen defines the relaxation time which is model-dependent and relates specifically to the analytic structure of the overlaps with the initial state~\cite{Cubero, Singularity}. For instance,  see Eq.~\eqref{exp}, for mass quenches in free theories the relaxation time can be calculated starting from the second-order in perturbation theory.  To test the robustness of the first-order result in an interacting theory, we performed in~\cite{SM} additional numerical simulations. They indicate that along the $E_8$ line, see Fig.~\ref{fig_quenches}, the order parameter $\sigma^x$ does not relax and the entanglement entropies still shows long-living oscillations also when the quench parameter $\delta h_x/h_x$ is of order one. Remarkably then, even after a large longitudinal field quench, the late-time dynamics continues to be qualitatively captured by first-order perturbation theory.

The formalism in this Letter allows calculating entanglement entropies in the scaling limit for any interacting massive theory   without fine-tuning of the initial state.  A priori this includes non-integrable models, even if the development of a perturbation series will generally be more challenging. It could be  adapted to quench protocols in absence of translation invariance~\cite{Delfino_pert} and we  believe that it   will be useful for other one-dimensional systems~\cite{PhysRevA.95.023621, C3, C4} that show similar long-living oscillations.

{\em Acknowledgements.---} We are grateful to A. Cubero, A. Delfino and G. Tak\'acs for numerous comments on the first version of this paper.  OACA and IMSZ  gratefully acknowledge support from EPSRC through the standard proposal ``Entanglement Measures, Twist Fields, and Partition Functions in Quantum Field Theory" under reference number EP/P006108/1. They are also grateful to the International Institute of Physics in Natal (Brazil) for support during the workshop ``Emergent Hydrodynamics in Low-Dimensional Quantum Systems" held in May 2019 during which discussions relating to this work took place. 
The work of IMSZ was supported by the grant ``Exact Results in Gauge and String Theories" from the Knut and Alice Wallenberg foundation; JV is supported by the Brazilian Ministeries MEC and MCTC and the Italian Ministery MIUR under the grant PRIN 2017  ``Low-dimensional quantum systems: theory, experiments and simulations".  Part of ML's work was carried out at the International Institute of Physics in Natal (Brazil), where his research was supported by the Brazilian Ministeries MEC and MCTC. ML also acknowledges support provided from the National Research,
Development and Innovation Office of Hungary, Project no. 132118 financed under the PD\_19 funding scheme. We thank  the High-Performance Computing Center (NPAD) at UFRN for providing computational resources.


\begin{thebibliography}{10}

\bibitem{ETH2}
J.~M. Deutsch,
\newblock Quantum statistical mechanics in a closed system,
\newblock Phys. Rev. A {\bf 43}, 2046--2049 (1991).

\bibitem{ETH3}
M.~Srednicki,
\newblock Chaos and quantum thermalization,
\newblock Phys. Rev. E {\bf 50}, 888--901 (1994).

\bibitem{kinoshita}
T.~Kinoshita, T.~Wenger, and D.~Weiss,
\newblock {A Quantum Newton's Cradle},
\newblock Nature {\bf 440}, 900 (2006).

\bibitem{Schmidt_07}
S.~Hofferberth, I.~Lesanovsky, B.~Fischer, T.~Schumm, and J.~Schmiedmayer,
\newblock Non-equilibrium coherence dynamics in one-dimensional Bose gases,
\newblock Nature {\bf 449}, 324--327 (2007).

\bibitem{Gring1318}
M.~Gring, M.~Kuhnert, T.~Langen, T.~Kitagawa, B.~Rauer, M.~Schreitl, I.~Mazets,
  D.~A. Smith, E.~Demler, and J.~Schmiedmayer,
\newblock Relaxation and Prethermalization in an Isolated Quantum System,
\newblock Science {\bf 337}(6100), 1318--1322 (2012).

\bibitem{Langen207}
T.~Langen, S.~Erne, R.~Geiger, B.~Rauer, T.~Schweigler, M.~Kuhnert,
  W.~Rohringer, I.~E. Mazets, T.~Gasenzer, and J.~Schmiedmayer,
\newblock Experimental observation of a generalized Gibbs ensemble,
\newblock Science {\bf 348}(6231), 207--211 (2015).

\bibitem{EEquench}
P.~Calabrese and J.~L. Cardy,
\newblock {Evolution of entanglement entropy in one-dimensional systems},
\newblock J. Stat. Mech. {\bf 0504}, P04010 (2005).

\bibitem{quench}
P.~Calabrese and J.~L. Cardy,
\newblock {Time-dependence of correlation functions following a quantum
  quench},
\newblock Phys. Rev. Lett. {\bf 96}, 136801 (2006).

\bibitem{CEF}
P.~Calabrese, F.~H.~L. Essler, and M.~Fagotti,
\newblock {Quantum Quench in the Transverse Field Ising Chain},
\newblock Phys. Rev. Lett. {\bf 106}(22), 227203 (2011).

\bibitem{FE}
M.~Fagotti and F.~H.~L. Essler,
\newblock Reduced density matrix after a quantum quench,
\newblock Phys. Rev. {\bf B87}, 245107 (2013).

\bibitem{Essler2016}
F.~H.~L. Essler and M.~Fagotti,
\newblock Quench dynamics and relaxation in isolated integrable quantum spin
  chains,
\newblock J. Stat. Mech. {\bf 2016}(6), 064002 (2016).

\bibitem{Mori_2018}
T.~Mori, T.~N. Ikeda, E.~Kaminishi, and M.~Ueda,
\newblock Thermalization and prethermalization in isolated quantum systems: a
  theoretical overview,
\newblock J. Phys. {\bf B51}(11), 112001 (2018).

\bibitem{AC}
V.~Alba and P.~Calabrese,
\newblock Entanglement and thermodynamics after a quantum quench in integrable
  systems,
\newblock PNAS {\bf 114}(30), 7947--7951 (2017).

\bibitem{FC}
M.~Fagotti and P.~Calabrese,
\newblock Evolution of entanglement entropy following a quantum quench:
  Analytic results for the $XY$ chain in a transverse magnetic field,
\newblock Phys. Rev. {\bf A78}, 010306 (2008).

\bibitem{Nahum_2016}
A.~Nahum, J.~Ruhman, S.~Vijay, and J.~Haah,
\newblock Quantum Entanglement Growth under Random Unitary Dynamics,
\newblock Phys. Rev. {\bf X7}, 031016 (2017).

\bibitem{PhysRevX.8.041019}
A.~Chan, A.~De~Luca, and J.~T. Chalker,
\newblock Solution of a Minimal Model for Many-Body Quantum Chaos,
\newblock Phys. Rev. {\bf X8}, 041019 (2018).

\bibitem{PhysRevX.8.021062}
P.~Kos, M.~Ljubotina, and T.~Prosen,
\newblock Many-Body Quantum Chaos: Analytic Connection to Random Matrix Theory,
\newblock Phys. Rev. {\bf X8}, 021062 (2018).

\bibitem{PhysRevX.9.021033}
B.~Bertini, P.~Kos, and T.~Prosen,
\newblock Entanglement Spreading in a Minimal Model of Maximal Many-Body
  Quantum Chaos,
\newblock Phys. Rev. {\bf X9}, 021033 (2019).


\bibitem{PhysRevLett.106.050405}
M.~C. Ba\~nuls, J.~I. Cirac, and M.~B. Hastings,
\newblock Strong and Weak Thermalization of Infinite Nonintegrable Quantum
  Systems,
\newblock Phys. Rev. Lett. {\bf 106}, 050405 (2011).

\bibitem{scar_Exp}
H.~Bernien, S.~Schwartz, A.~Keesling, H.~Levine, A.~Omran, H.~Pichler, S.~Choi,
  A.~S. Zibrov, M.~Endres, M.~Greiner, V.~Vuletic, and M.~Lukin,
\newblock Probing many-body dynamics on a 51-atom quantum simulator,
\newblock Nature {\bf 551} (7682)  (2017).

\bibitem{Abanin}
C.~Turner, A.~Michailidis, D.~Abanin, M.~Serbyn, and Z.~Papic,
\newblock Weak ergodicity breaking from quantum many-body scars,
\newblock Nature Physics {\bf 14}, 745--749 (2018).

\bibitem{Gabor1}
M.~Kormos, M.~Collura, G.~Tak\'acs, and P.~Calabrese,
\newblock {Real time confinement following a quantum quench to a non-integrable
  model},
\newblock Nature Physics. {\bf 13}, 246--249 (2017).

\bibitem{RAKOVSZKY2016805}
T.~Rakovszky, M.~Mesty\'an, M.~Collura, M.~Kormos, and G.~Tak\'acs,
\newblock Hamiltonian truncation approach to quenches in the Ising field
  theory,
\newblock Nucl. Phys. {\bf B911}, 805--845 (2016).

\bibitem{10.21468/SciPostPhys.5.3.027}
K.~Hodsagi, M.~Kormos, and G.~Takacs,
\newblock {Quench dynamics of the Ising field theory in a magnetic field},
\newblock SciPost Phys. {\bf 5}, 27 (2018).

\bibitem{Konik1}
A.~J.~A. James, R.~M. Konik, and N.~J. Robinson,
\newblock Nonthermal States Arising from Confinement in One and Two Dimensions,
\newblock Phys. Rev. Lett. {\bf 122}, 130603 (2019).

\bibitem{Konik2}
N.~J. Robinson, A.~J.~A. James, and R.~M. Konik,
\newblock Signatures of rare states and thermalization in a theory with
  confinement,
\newblock Phys. Rev. {\bf B99}, 195108 (2019).

\bibitem{C1}
T.~Chanda, J.~Zakrzewski, M.~Lewenstein, and L.~Tagliacozzo,
\newblock {Confinement and lack of thermalization after quenches in the bosonic
  Schwinger model},
\newblock 1909.12657  (2019).

\bibitem{C3}
A.~Lerose, F.~M. Surace, P.~Mazza, G.~Perfetto, M.~Collura, and A.~Gambassi,
\newblock Quasilocalized dynamics from confinement of quantum excitations,
\newblock 1911.07877  (2019).

\bibitem{Cubero2}
A.~Cubero and N.~Robinson,
\newblock {Lack of thermalization in (1+1)-d QCD at large Nc},
\newblock arXiv:1908.00270  (2020).

\bibitem{Magnifico}
G.~Magnifico, M.~Dalmonte, P.~Facchi, S.~Pascazio, F.~Pepe, and E.~Ercolessi,
\newblock Real Time Dynamics and Confinement in the $\mathbb{Z}_{n}$
  Schwinger-Weyl lattice model for 1+1 QED,
\newblock arXiv:1909.04821  (2019).

\bibitem{C2}
S.~Pai and M.~Pretko,
\newblock Fractons from confinement,
\newblock 1909.12306  (2019).

\bibitem{TC}
M.~Medenjak, B.~Buca, and D.~Jaksch,
\newblock The isolated Heisenberg magnet as a quantum time crystal,
\newblock arXiv:1905.08266  (2019).

\bibitem{C4}
F.~Liu, R.~Lundgren, P.~Titum, G.~Pagano, J.~Zhang, C.~Monroe, and A.~V.
  Gorshkov,
\newblock Confined Quasiparticle Dynamics in Long-Range Interacting Quantum
  Spin Chains,
\newblock Phys. Rev. Lett. {\bf 122}, 150601 (2019).

\bibitem{korepinbook}
V.~Korepin, N.~Bogoliugov, and A.~Izergin,
\newblock Quantum inverse scattering method and correlation functions,
\newblock Cambridge University Press, Cambridge  (1993).

\bibitem{PQ1}
G.~Delfino,
\newblock Quantum quenches with integrable pre-quench dynamics,
\newblock J. Phys. {\bf A47}(40), 402001 (2014).

\bibitem{PQ2}
G.~Delfino and J.~Viti,
\newblock {On the theory of quantum quenches in near-critical systems},
\newblock J. Phys. {\bf A50}(8), 084004 (2017).

\bibitem{Calabrese:2004eu}
P.~Calabrese and J.~L. Cardy,
\newblock Entanglement entropy and quantum field theory,
\newblock J. Stat. Mech. {\bf 0406}, P002 (2004).

\bibitem{entropy}
J.~L. Cardy, O.~A. Castro-Alvaredo, and B.~Doyon,
\newblock Form factors of branch-point twist fields in quantum integrable
  models and entanglement entropy,
\newblock J. Stat. Phys. {\bf 130}, 129--168 (2008).

\bibitem{review}
O.~A. Castro-Alvaredo and B.~Doyon,
\newblock {Bi-partite entanglement entropy in massive 1+1-dimensional quantum
  field theories},
\newblock J. Phys. {\bf A42}, 504006 (2009).

\bibitem{ourIsing}
O.~A. Castro-Alvaredo, M.~Lencs\'es, I.~M. Sz\'ecs\'enyi, and J.~Viti,
\newblock {Entanglement Dynamics after a Quench in Ising Field Theory: A Branch
  Point Twist Field Approach},
\newblock JHEP {\bf 2019}, 79 (2019).

\bibitem{BPZ}
A.~A. Belavin, A.~M. Polyakov, and A.~B. Zamolodchikov,
\newblock {Infinite Conformal Symmetry in Two-Dimensional Quantum Field
  Theory},
\newblock Nucl. Phys. {\bf B241}, 333--380 (1984).

\bibitem{Zamolodchikov:1989fp}
A.~B. Zamolodchikov,
\newblock {Integrals of Motion and S Matrix of the (Scaled) $T=T_c$ Ising Model
  with Magnetic Field},
\newblock Int. J. Mod. Phys. {\bf A4}, 4235 (1989).

\bibitem{Delfino:2003yr}
G.~Delfino,
\newblock Integrable field theory and critical phenomena: The Ising model in a
  magnetic field,
\newblock J. Phys. {\bf A37}, R45 (2004).

\bibitem{qft}
G.~Mussardo,
\newblock Statistical Field Theory: An introduction to exactly solved models in
  Statistical Physics (Second Ed.),
\newblock OUP  (2020).

\bibitem{PIM}
G.~Delfino, G.~Mussardo, and P.~Simonetti,
\newblock {Nonintegrable quantum field theories as perturbations of certain
  integrable models},
\newblock Nucl. Phys. {\bf B473}, 469--508 (1996).

\bibitem{Overlap1}
K.~Hodsagi, M.~Kormos, and G.~Takacs,
\newblock {Perturbative post-quench overlaps in quantum field theory},
\newblock JHEP {\bf 2019} (2019).


\bibitem{foot}
If $|p_1,\dots,p_{n}\rangle_{\text{post}}^{\text{out}}$ is a $n$-particle out state of $H_{\text{post}}$, the Lorentz invariant integration measure in $1+1$ dimensions is $\sum_{\alpha}:=\sum_{n=0}^{\infty}\int_{p_1<p_2<\dots<p_n}\prod_{j=1}^n\frac{dp_j}{2\pi\tilde{e}(p_j)}$.

\bibitem{smirnovbook}
F.~Smirnov,
\newblock Form factors in completely integrable models of quantum field theory,
\newblock Adv. Series in Math. Phys. {\bf 14}, World Scientific, Singapore
  (1992).

 \bibitem{Foot2} 
If $|p_1,\dots,p_n\rangle$ is a $n$-particle eigenstate of the pre-quench Hamiltonain $H$, the symmetrized Lorentz invariant integration measure in the text  is $\sum_{\alpha}':=\sum_{n=0}^{\infty}\frac{1}{n!}\int_{\mathbb {R}^n}\prod_{j=1}^n\frac{dp_j}{2\pi e(p_j)}$.   
  
\bibitem{orbifold}
L.~Dixon, D.~Friedan, E.~Martinec, and S.~Shenker,
\newblock The conformal field theory of orbifolds,
\newblock Nucl. Phys. {\bf B282}, 13--73 (1987).

\bibitem{Bouwknegt}
P.~Bouwknegt,
\newblock {Coset construction for winding subalgebras and applications},
\newblock q-alg/9610013 .

\bibitem{Borisov}
L.~Borisov, M.~B. Halpern, and C.~Schweigert,
\newblock {Systematic approach to cyclic orbifolds},
\newblock Int. J. Mod. Phys. {\bf A13}, 125--168 (1998).

\bibitem{Zamolodchikov:1989zs}
A.~B. Zamolodchikov,
\newblock {Integrable field theory from conformal field theory},
\newblock Adv. Stud. Pure Math. {\bf 19}, 641--674 (1989).

\bibitem{Coldea177}
R.~Coldea, D.~A. Tennant, E.~M. Wheeler, E.~Wawrzynska, D.~Prabhakaran,
  M.~Telling, K.~Habicht, P.~Smeibidl, and K.~Kiefer,
\newblock Quantum Criticality in an Ising Chain: Experimental Evidence for
  Emergent $E_8$ Symmetry,
\newblock Science {\bf 327}(5962), 177--180 (2010).

\bibitem{PhysRevB.83.020407}
J.~A. Kj\"all, F.~Pollmann, and J.~E. Moore,
\newblock Bound states and ${E}_{8}$ symmetry effects in perturbed quantum
  Ising chains,
\newblock Phys. Rev. {\bf B83}, 020407 (2011).

\bibitem{SM}
See Supplemental Material [url] for details about the analytic calculation of the entropies and the numerical study in the scaling limit. The Supplemental Material also includes Refs. \cite{fateev, DS, Pollmann_notes, FOREST1990105, CalCarPesUnusual, 2010JSMTE..04..023C} below.

\bibitem{fateev}
V.~Fateev,
\newblock The exact relations between the coupling constants and the masses of
  particles for the integrable perturbed conformal field theories,
\newblock Phys. Lett. {\bf B324}(1), 45--51 (1994).

\bibitem{DS}
G.~Delfino and P.~Simonetti,
\newblock {Correlation functions in the two-dimensional Ising model in a
  magnetic field at $T = T_c$},
\newblock Phys. Lett. {\bf B383}, 450--456 (1996).


\bibitem{Pollmann_notes}
F.~Pollmann,
\newblock Efficient Numerical Simulations Using Matrix-Product States, 2016.

\bibitem{FOREST1990105}
E.~Forest and R.~D. Ruth,
\newblock Fourth-order symplectic integration,
\newblock Physica {\bf 43D}(1), 105 -- 117 (1990).

\bibitem{CalCarPesUnusual}
P.~Calabrese, J.~Cardy, and I.~Peschel,
\newblock {Corrections to scaling for block entanglement in massive
  spin-chains},
\newblock J. Stat. Mech. {\bf 1009}, P09003 (2010).

\bibitem{2010JSMTE..04..023C}
J.~{Cardy} and P.~{Calabrese},
\newblock {Unusual corrections to scaling in entanglement entropy},
\newblock J. Stat. Mech. {\bf 2010}(4), 04023 (2010).


\bibitem{Fateev:1997yg}
V.~Fateev, S.~L. Lukyanov, A.~B. Zamolodchikov, and A.~B. Zamolodchikov,
\newblock {Expectation values of local fields in Bullough-Dodd model and
  integrable perturbed conformal field theories},
\newblock Nucl. Phys. {\bf B516}, 652--674 (1998).

\bibitem{DM}
G.~Delfino and G.~Mussardo,
\newblock {The Spin spin correlation function in the two-dimensional Ising
  model in a magnetic field at $T =T_c$},
\newblock Nucl. Phys. {\bf B455}, 724--758 (1995).

\bibitem{E8toda}
O.~A. Castro-Alvaredo,
\newblock {Massive Corrections to Entanglement in Minimal $E_8$ Toda Field
  Theory},
\newblock SciPost Phys. {\bf 2}(1), 008 (2017).

\bibitem{VidaliTEBD1}
G.~Vidal,
\newblock Efficient Simulation of One-Dimensional Quantum Many-Body Systems,
\newblock Phys. Rev. Lett. {\bf 93}, 040502 (2004).

\bibitem{VidaliTEBD2}
G.~Vidal,
\newblock Classical Simulation of Infinite-Size Quantum Lattice Systems in One
  Spatial Dimension,
\newblock Phys. Rev. Lett. {\bf 98}, 070201 (2007).

\bibitem{Henkel_1989}
M.~Henkel and H.~Saleur,
\newblock The two-dimensional Ising model in the magnetic field: a numerical
  check of Zamolodchikov conjecture,
\newblock J. Phys. {\bf A22}(11), L513--L518 (1989).

\bibitem{SE}
D.~Schuricht and F.~H.~L. Essler,
\newblock {Dynamics in the Ising field theory after a quantum quench},
\newblock J. Stat. Mech. {\bf 1204}, P04017 (2012).

\bibitem{PhysRevA.95.023621}
C.-J. Lin and O.~I. Motrunich,
\newblock Quasiparticle explanation of the weak-thermalization regime under
  quench in a nonintegrable quantum spin chain,
\newblock Phys. Rev. {\bf A95}, 023621 (2017).


\bibitem{Cubero}
A.~Cubero and D.~Schuricht,
\newblock Quantum quench in the attractive regime of the sine-Gordon model,
\newblock J. Stat. Mech. {\bf 103106} (2017).

\bibitem{Singularity}
D.~Horvath, M.~Kormos, and G.~Takacs,
\newblock Overlap singularity and time evolution in integrable quantum field
  theory,
\newblock JHEP {\bf 2018} (2018).

\bibitem{Delfino_pert}
G.~Delfino,
\newblock Persistent oscillations after quantum quenches,
\newblock Nucl. Phys. {\bf B954} (2020) 115002.

\end{thebibliography}
\end{document}


{\begin{center}
{\Large \bf SUPPLEMENTARY MATERIAL} 
\end{center}}
The Supplementary Material is organized as follows.
In  Section~\ref{ff} we summarize the main analytical results that we have used in our Letter, particularly when comparing the predictions of quench perturbation theory \cite{PQ1,PQ2} to lattice numerical calculations in the scaling limit. 
These analytical results are principally the form factors of local fields in the theories under consideration. The fields involved here are the branch point twist field whose correlators are directly linked to measures of entanglement, and the field associated with the coupling whose sudden change generates the quench.

  In our Letter we have mainly discussed the longitudinal field quench (for critical transverse field). In the QFT setting,  this is equivalent to a  mass quench in the $E_8$ minimal Toda field theory. The mass  $m_{0,1}$ of the lightest particle in the model before the quench  (see Section~\ref{ff}) and the coupling constant $\lambda_2$ are related as
  \beq
  m_{0,1}= \kappa \,\lambda_2^{\frac{8}{15}}\,, \qquad \mathrm{with} \qquad \kappa= 4.40490858... \,,
  \label{mc}
  \eeq
  where the constant $\kappa$ has an analytical expression in terms of $\Gamma$-functions first found in \cite{fateev}. Therefore a change of the longitudinal field $h_x$ proportional to $\lambda_2$ is equivalent to a change of the masses of the eight particles in the spectrum, which are all multiples of $m_{0,1}$.
   As shown in Fig.~1 and equation (2) of the Letter, in the QFT,  the perturbing field is the spin field $\sigma(x,t)$.  We analyze such a quench in Section 1 of the supplementary material. 

In Section~\ref{exp},  we also present a complete proof of the exponentiation of the one-point function of the branch point twist field in the case of a transverse field quench (for zero longitudinal field).  That is equivalent to a mass quench in Ising field theory since now the fermion mass $m_0$ is proportional to $|h_z-1|$, where $h_z$ is the transverse field. Our proof complements the detailed study presented in \cite{ourIsing}. In this case the perturbing field is the energy field $\varepsilon(x,t)$ as also shown in Fig.~1 of the Letter.

In Section~\ref{num} we provide a description of the numerical techniques, with additional results not included in the Letter. Finally in Section~\ref{bfo} we comment on the robustness of  entanglement oscillations at higher order in perturbation theory.

\medskip

\section{Longitudinal Field Quench: Mass Quench in $E_8$ Minimal Toda Field Theory}
\label{ff}
Let us start by fixing the basic definitions and notations for the form factors that we have employed in the present work. Although in our Letter we have written formulae involving form factors of an arbitrary state $|\alpha\ket^{\rm{out/in}}$, in practise the only form factors that  have been analytically evaluated are those associated with one- and two-particle states. We will therefore only review those here. 
 In a 1+1-dimensional QFT we define the one- and the two-particle form factors of a local, spinless field $\mathcal{O}$  as the matrix elements
\beq
\label{formf}
F_{a_1}^{\mathcal{O}}:=\bra \Omega| \mathcal{O}(0,0)|\theta_1 \ket_{a_1}\qquad \mathrm{and}\qquad F_{a_1 a_2}^{\mathcal{O}}(\theta_1-\theta_2):= \bra \Omega | \mathcal{O}(0,0)|\theta_1 \theta_2 \ket_{a_1 a_2}\,,
\eeq
where $a_1, a_2$ are particle quantum numbers, $\theta_1, \theta_2$ are rapidities, in terms of which---and with a slight abuse of notations--- the energy and momentum are given by $e_a(\theta)=m_{0,a}\cosh\theta$ and $p_a(\theta)=m_{0,a}\sinh\theta$, where $m_{0,a}$ is the mass of a particle of type $a$ in the pre-quench theory.  The state $|\Omega\rangle$ is the pre-quench vacuum.

In general, $|\theta_1\ldots \theta_k\ket_{a_1\ldots a_k}\,,$ is an asymptotic $in$-state of $k$ particles and $|\Omega\ket$ is the ground state. For spinless fields in relativistic QFT the one-particle form factors are rapidity-independent whereas the two-particle form factors depend only upon rapidity differences, hence our notation.  It is also useful to introduce the normalized one-particle form factor as
\begin{equation}
 \hat{F}_{a_1}^{\mathcal{O}}:=\frac{\bra \Omega| \mathcal{O}(0,0)|\theta_1 \ket_{a_1}}{\langle\Omega|\mathcal{O}(0,0)|\Omega\rangle};
\end{equation}
for the spin field and the twist field we will use the shorthand notation $\langle\Omega|\sigma(0,0)|\Omega\rangle\equiv\bar{\sigma}$ and $\langle\Omega|\mathcal{T}_n(0,0)|\Omega\rangle\equiv\tau_n$.

The $E_8$ minimal Toda field theory  is an integrable model with diagonal scattering matrix and an eight particle spectrum. They were both first given in the seminal papers \cite{Zamolodchikov:1989zs,Zamolodchikov:1989fp}. The mass spectrum takes the form 
\beq
r_2=2\cos\frac{\pi}{5}\,, \quad r_3=2\cos\frac{\pi}{30}\,,\quad r_4=2r_2 \cos\frac{7\pi}{30}\,, \quad r_5=2r_2 \cos\frac{2\pi}{15}\,,  \nonumber
\eeq
\beq
 r_6=r_2r_3\,,\quad r_7=r_2r_4\,,\quad r_8=r_2r_5\,,
\eeq
where $r_i:=m_{0,i}/m_{0,1}$ (hence, $r_1=1$). It should be noticed that the ratios $r_i$ are the same for both the pre-quench and post-quench theories as a consequence of universality of the scaling limit.

\subsection{One-Particle Form Factors}
As we have seen in equation (9) of the Letter the calculation of the entanglement entropies is based upon the knowledge of the one-particle form factors of the branch point twist field and the spin field. So far the only computation of the twist field form factors in this theory was performed in \cite{OToda}. There, explicit values of the form factors of the first four lightest particles for $n=2$ were given, whereas for $2<n\leq 4$ the values of $\hat{F}_a^{\TT_n}$ were presented graphically. However, while carrying out detailed comparison with the lattice results in the scaling limit, we realized that it was critical to have a more accurate evaluation of the quantities $\hat{F}_a^{\TT_n}$  with $n=2,3,4$. The values given in \cite{OToda} were dependent on the asymptotic behaviour of certain two-particle form factors, and we have realized that this asymptotic value was systematically underestimated in \cite{OToda} .  In Table~\ref{tab1} we present the values of $\hat{F}_a^{\TT_n}$  with $n=2,3,4$  as newly obtained using a different technique (we will present further details in subsection 1.2). For $n=2$ they differ from those presented in \cite{OToda} by between 10\% and 20\% (depending on the particle). This has allowed us to reach a much improved matching with numerical values. 

\begin{table}
\[
\begin{array}{|c||c|c|c|}
\hline n & 2 & 3 & 4\\
\hline\hline \hat{F}_{1}^{\mathcal{T}_n} & -0.17124900374494678 & -0.1996259878353373 & -0.20930848250173\\
\hline \hat{F}_{2}^{\mathcal{T}_n} & 0.07005900535572894& 0.08788104227633028& 0.094313951168679\\
\hline \hat{F}_{3}^{\mathcal{T}_n} & -0.03440203483936546 & -0.04473113820315836& -0.048562882373633\\
\hline \hat{F}_{4}^{\mathcal{T}_n} & 0.023657419055677746 & 0.031870010789866426& 0.0349996171170825
\\\hline \end{array}
\]
\caption{Newly evaluated (normalized) one-particle form factors of the branch point twist field for the four lightest particles in the spectrum. The hat in $\hat{F}_i^{\TT_n}$ indicates normalization by the vacuum expectation value of the branch point twist field.}
\label{tab1}
\end{table}

As we can see, equation (9) in the Letter also requires the knowledge of one-particle form factors of the spin field. These were computed in \cite{DM,DS}. Their (normalized) values  are given  in Table~2.
\begin{table}[htbp]
\begin{center}
\begin{tabular}{|c|c|c|c|}
\hline
$\hat{F}_1^{\sigma}$&$\hat{F}_2^{\sigma}$&$\hat{F}_3^{\sigma}$&$\hat{F}_4^{\sigma}$\\\hline
\hline
-0.64090211&0.33867436&-0.18662854&0.14277176\\\hline
\hline
$\hat{F}_5^{\sigma}$&$\hat{F}_6^{\sigma}$&$\hat{F}_7^{\sigma}$&$\hat{F}_8^{\sigma}$\\\hline
\hline
0.06032607&-0.04338937&0.01642569&-0.00303607\\\hline
\end{tabular}
\end{center}
\caption{Normalized one-particle form factors of the spin field $\hat{F}_i^{\sigma}=F_i^{\sigma} / \bar{\sigma} $, where $\bar{\sigma}$ is the expectation value of the spin field.}
\end{table}

In order to compute the von Neumann entropy we also need the values of
\beq
g_a:=\lim_{n\rightarrow 1} \frac{\hat{F}_a^{\TT_n}}{1-n}\,.
\label{ga}
\eeq
\begin{table}
\[
\begin{array}{|c||c|}
\hline  g_1 & -0.4971505471133315\\
\hline  g_2 & 0.10034674000675149\\
\hline  g_3& -0.03659195487267996\\
\hline g_4 & 0.01914945194919403
\\\hline \end{array}
\]
\caption{The functions (\ref{ga}) for the four lightest particles in the spectrum.}
\end{table}
It is well-known that the functions $\hat{F}_a^{\TT_n}$ tend to zero for $n\rightarrow 1$. However, the precise asymptotics was not investigated in \cite{OToda}. This asymptotics can be studied by analysing the consistency equations that these form factors must satisfy and which were given in \cite{OToda}. These give rise to a Taylor expansion in powers of $n-1$, starting with power one. Once more, the derivation relies heavily on properties of the functions that enter the two particle form factors. We will briefly discuss these below. Table 3 gives the values of the first four functions $g_a$. 

\subsection{Two-Particle Form Factors}

In \cite{OToda} the two-particle form factors $F_{11}^{\TT_n}(\theta)$ and $F_{12}^{\TT_n}(\theta)$ were also computed. Here we are referring always to particles in the same copy so we omit copy numbers. 
They are given by
\beqa 
F_{11}^{\TT_n}(\theta)=\frac{\tau_n Q^{\TT_n}_{11}(\theta)}
{2n K_{11}(\theta;n)\prod\limits_{\alpha=\frac{2}{3},\frac{2}{5},\frac{1}{15}}B_\alpha(\theta;n)}\frac{f_{11}(\theta;n)}{f_{11}(i\pi;n)}\,, \label{genff2}
\eeqa
and
\beqa 
F_{12}^{\TT_n}(\theta)=\frac{\tau _n Q^{\TT_n}_{12}(\theta)}
{2n \prod\limits_{\alpha=\frac{4}{5},\frac{3}{5},\frac{7}{15},\frac{4}{15}} B_\alpha(\theta;n)}\frac{f_{12}(\theta;n)}{f_{12}(i\pi;n)}\,, \label{genff3}
\eeqa 
with
\beq
f_{11}(\theta;n)= -i \sinh\frac{\theta}{2n}\exp\left[2 \int_{0}^{\infty}  \frac{\mathrm{d}t}{t} \frac{\cosh\frac{t}{10}+\cosh\frac{t}{6}+\cosh\frac{13t}{30}}{ \cosh\frac{t}{2}\sinh(n t)}  \sin^{2}\left(\frac{i t}{2}\left(n+\frac{i\theta}{\pi}\right)\right)\right]\,,\label{f11}
\eeq 
and
\beq
f_{12}(\theta;n)=\exp\left[2 \int_{0}^{\infty}  \frac{\mathrm{d}t}{t} \frac{\cosh\frac{t}{10}+\cosh\frac{3t}{10}+\cosh\frac{t}{30}+\cosh\frac{7t}{30}}{ \cosh\frac{t}{2}\sinh(n t)}  \sin^{2}\left(\frac{i t}{2}\left(n+\frac{i\theta}{\pi}\right)\right)\right]\,,\label{f12}
\eeq 
where
\beq 
K_{11}(\theta;n)=\frac{\sinh\left(\frac{i\pi-\theta}{2n}\right)\sinh\left(\frac{i\pi+\theta}{2n}\right)}{\sin\frac{\pi}{n}},
\eeq 
\beq
B_{\alpha}(\theta;n)=\sinh\left(\frac{i \pi \alpha-\theta}{2n}\right)\sinh\left(\frac{i\pi\alpha+\theta}{2n} \right),
\eeq 
$\tau_n$ is the vacuum expectation value of the branch point twist field and the functions $Q_{ij}^{\TT_n}(\theta)$ have the general structure
\beq 
Q^{\TT_n}_{11}(\theta)=A_{11}(n)+B_{11}(n) \cosh\frac{\theta}{n}+C_{11}(n) \cosh^2\frac{\theta}{n},
\label{q1}
\eeq 
and 
\beq 
Q^{\TT_n}_{12}(\theta)=A_{12}(n)+B_{12}(n) \cosh\frac{\theta}{n}+C_{12}(n) \cosh^2\frac{\theta}{n}\,,
\label{q2}
\eeq 
with coefficients that can be determined for each value of $n$. They have also been re-evaluated with greater precision for this work and are listed in Table~4. 
\begin{table}[h!]
\[
\begin{array}{|c||c|c|c|}
\hline n & 2 & 3 & 4\\
\hline A_{11}(n) & 0.05028656966443226 & 0.008222663493673649 & 0.003094920413703075\\
\hline B_{11}(n) & 0.0011995725313121055 & -0.005436476679120888& -0.004685055209534106\\
\hline C_{11}(n) & 0.009415788449439522 & 0.0053536637424471565 & 0.0032612776790863058\\
\hline A_{12}(n) & -0.03893395394244009 & -0.0052884492808026595 & -0.0014212064608825764\\
\hline B_{12}(n) & -0.0006333785909299339 & 0.0024712680272948595 & 0.0017331405976600545\\
\hline C_{12}(n) & -0.004484918093780377& -0.002327262587055512 & -0.0011767365122042212
\\\hline \end{array}
\]
\caption{The coefficients of the functions $Q^{\TT_n}_{11}(\theta)$ and $Q^{\TT_n}_{12}(\theta)$.}
\end{table}

At the heart of these improved values is the improved evaluation of the the leading asymptotics of the functions $f_{11}(\theta,n)$, $f_{12}(\theta,n)$ for $\theta\rightarrow \infty$. 
The understanding of this asymptotic plays an important role in fixing the one-particle form factors because the two-particle form factors are expected to satisfy the clustering property in momentum space, namely 
\beq
\lim_{\theta \rightarrow \infty} F_{a_1 a_2}^{\TT_n}(\theta)\,\propto\, F_{a_1}^{\TT_n}  F_{a_2}^{\TT_n}\,.
\eeq
For $a_1=a_2=1$ and $a_1=1, a_2=2$ this gives two of the conditions that were employed in \cite{OToda} to fix the one-particle form factors. 
It turns out that both functions $f_{11}(\theta,n)$ and $f_{12}(\theta,n)$ can be expressed as products of the following blocks:
\begin{eqnarray}
f\left(\theta,\alpha;n\right) = \exp\left\{ 2\int_{0}^{\infty}\frac{\mathrm{d}t}{t}\frac{\cosh\left[t\left(\alpha-\frac{1}{2}\right)\right]}{\cosh\left(\frac{t}{2}\right)}\frac{\sin^{2}\left(\frac{t\left[i\pi n-\theta\right]}{2\pi}\right)}{\sinh\left(nt\right)}\right\} \,,\label{eq:fmin_n_alpha}
\end{eqnarray}
for particular choices of $\alpha$. In fact
\beq
f_{11}(\theta,n)=f(\theta,0;n) f(\theta,\frac{2}{3};n)f(\theta,\frac{2}{5};n)f(\theta,\frac{1}{15};n)\,,
\eeq
\beq
f_{12}(\theta,n)=f(\theta,\frac{4}{5};n) f(\theta,\frac{3}{5};n)f(\theta,\frac{7}{15};n)f(\theta,\frac{4}{15};n)\,.
\eeq
A natural simplification of the formula (\ref{eq:fmin_n_alpha}) that helps to extract the asymptotics is to pull out a factor $f_0(\theta,n):=f(\theta,0;n)$ using the integral representation
\begin{eqnarray}
f\left(\theta,\alpha;n\right) & = & f_0\left(\theta;n\right)\exp\left\{ 2\int_{0}^{\infty}\frac{\mathrm{d}t}{t}\left[\frac{\cosh\left[t\left(\alpha-\frac{1}{2}\right)\right]}{\cosh\left(\frac{t}{2}\right)}-1\right]\frac{\sin^{2}\left(\frac{t\left[i\pi n-\theta\right]}{2\pi}\right)}{\sinh\left(nt\right)}\right\} \nonumber \\
 & = & f_0\left(\theta;n\right)\exp\left\{ -4\int_{0}^{\infty}\frac{\mathrm{d}t}{t}\frac{\sinh\left(\frac{t}{2}\alpha\right)\sinh\left(\frac{t}{2}(1-\alpha)\right)}{\cosh\left(\frac{t}{2}\right)}\frac{\sin^{2}\left(\frac{t\left[i\pi n-\theta\right]}{2\pi}\right)}{\sinh\left(nt\right)}\right\} \\
& = & f_0\left(\theta;n\right)\mathcal{N}\left(\alpha;n\right)\exp\left\{ 2\int_{0}^{\infty}\frac{\mathrm{d}t}{t}\frac{\sinh\left(\frac{t}{2}\alpha\right)\sinh\left(\frac{t}{2}(1-\alpha)\right)}{\cosh\left(\frac{t}{2}\right)}\frac{\cos\left(\frac{t\left[i\pi n-\theta\right]}{\pi}\right)}{\sinh\left(nt\right)}\right\}\,, \nonumber
\end{eqnarray}
with 
\begin{eqnarray}
\mathcal{N}\left(\alpha;n\right)  =  \exp\left\{ -2\int_{0}^{\infty}\frac{\mathrm{d}t}{t}\frac{\sinh\left(\frac{t}{2}\alpha\right)\sinh\left(\frac{t}{2}(1-\alpha)\right)}{\cosh\left(\frac{t}{2}\right)\sinh\left(nt\right)}\right\} \,.
\end{eqnarray}
 For large $\theta$, one can then show that
\begin{eqnarray}
\lim_{\theta\to\infty}f\left(\theta,\alpha;n\right)  =  \mathcal{N}\left(\alpha;n\right)\lim_{\theta\to\infty}f_{0}\left(\theta;n\right)\sim\frac{\mathcal{N}\left(\alpha;n\right)}{2i}e^{\frac{\theta}{2n}}\,.
\label{beh}
\end{eqnarray}
It is easy to show that $\mathcal{N}\left(\alpha;n\right)$ are convergent integrals, but the remaining integral
\begin{eqnarray}
\tilde{I}\left(\theta,\alpha;n\right) & = & 2\int_{0}^{\infty}\frac{\mathrm{d}t}{t}\frac{\sinh\left(\frac{t}{2}\alpha\right)\sinh\left(\frac{t}{2}(1-\alpha)\right)}{\cosh\left(\frac{t}{2}\right)}\frac{\cos\left(\frac{t\left[i\pi n-\theta\right]}{\pi}\right)}{\sinh\left(nt\right)\nonumber}\\
 & = & 4\int_{0}^{\infty}\frac{\mathrm{d}t}{t}\frac{\sinh\left(\frac{t}{2}\alpha\right)\sinh\left(\frac{t}{2}(1-\alpha)\right)\sinh\left(\frac{t}{2}\right)}{\sinh\left(t\right)}\frac{\cos\left(\frac{t\left[i\pi n-\theta\right]}{\pi}\right)}{\sinh\left(nt\right)}\,,
\end{eqnarray}
needs regularization. Our strategy is to express the $\sinh t$
in the denominator as 
\begin{eqnarray}
\frac{1}{\sinh t } & = & 2\sum_{k=0}^{N-1}e^{-(2k+1)t}+\frac{e^{-2Nt}}{\sinh t}\,,\label{eq:sinh_expansion}
\end{eqnarray}
and use the integral identity
\begin{eqnarray}
\int_{0}^{\infty}\frac{\mathrm{d}t}{t}\frac{\sinh\left(\beta t\right)\sinh\left(\gamma t\right)e^{-\mu t}}{\sinh\left(\delta t\right)} & = & \frac{1}{2}\log\omega\left(\beta,\gamma,\mu,\delta\right)\nonumber \\
 & = & \frac{1}{2}\log\left\{ \frac{\Gamma\left(\frac{\beta+\gamma+\mu+\delta}{2\delta}\right)\Gamma\left(\frac{-\beta-\gamma+\mu+\delta}{2\delta}\right)}{\Gamma\left(\frac{-\beta+\gamma+\mu+\delta}{2\delta}\right)\Gamma\left(\frac{\beta-\gamma+\mu+\delta}{2\delta}\right)}\right\} \,.\label{eq:sinh_int_formula}
\end{eqnarray}
The regularized integral then is 
\begin{eqnarray}
\tilde{I}\left(\theta,\alpha,N;n\right) & = & 2\int_{0}^{\infty}\frac{\mathrm{d}t}{t}\frac{\sinh\left(\frac{t}{2}\alpha\right)\sinh\left(\frac{t}{2}(1-\alpha)\right)}{\cosh\left(\frac{t}{2}\right)}\frac{\cos\left(\frac{t\left[i\pi n-\theta\right]}{\pi}\right)}{\sinh\left(nt\right)}e^{-2Nt} \label{eq:int_reg_1}\\
 &  & +\sum_{k=0}^{N-1}\log\frac{\omega\left(\frac{\alpha}{2},\frac{1-\alpha}{2},2k+\frac{1}{2}+n+i\frac{\theta}{\pi},n\right)\omega\left(\frac{\alpha}{2},\frac{1-\alpha}{2},2k+\frac{1}{2}-n-i\frac{\theta}{\pi},n\right)}{\omega\left(\frac{\alpha}{2},\frac{1-\alpha}{2},2k+\frac{3}{2}+n+i\frac{\theta}{\pi},n\right)\omega\left(\frac{\alpha}{2},\frac{1-\alpha}{2},2k+\frac{3}{2}-n-i\frac{\theta}{\pi},n\right)}\,. \nonumber 
\end{eqnarray}
The asymptotics of the $\Gamma$ function for large imaginary
values is 
\begin{eqnarray}
\lim_{y\to\infty}\Gamma(x+iy)  \sim  \sqrt{2\pi}y^{x-1/2+i y}e^{-i\pi /4 +i \pi x/2-i y -\pi y/2+\mathcal{O}(1/y) }\,,
\end{eqnarray}
hence
\begin{eqnarray}
\lim_{\theta\to\infty} \omega (\beta,\gamma,\mu +i \theta /\pi, n)=1\,.
\end{eqnarray}
Since the value of $\tilde{I}\left(\theta,\alpha,N;n\right)$ is independent of the value of $N$
\begin{eqnarray}
\lim_{\theta\to\infty}\tilde{I}\left(\theta,\alpha,N;n\right) =\lim_{\theta\to\infty}\tilde{I}\left(\theta,\alpha,\infty;n\right) =  0\,,
\end{eqnarray}
which then gives the behaviour (\ref{beh}). 

With the more accurate evaluation of the asymptotics, the solution of the consistency equations for the first four one-particle and first two two-particle form factors boils down of solving a cubic equation, whose appropriate solution is chosen by the property that it should vanish as $n$ approaches $1$. The expressions for the coefficients are cumbersome, hence we do not list them over here, only the numerical values for $n=2,3,4$ in Table~1 and Table~4. 

In order to evaluate the von Neumann entropy we also need the asymptotic values of the constants in equations (\ref{q1}) and (\ref{q2}) as $n\rightarrow 1$. This requires the values
\beq
\lim_{n\rightarrow 1} {(1-n) K_{11}(\theta;n)}= \pi \cosh^2\frac{\theta}{2}\,,
\eeq
and 
\beq
  B_\alpha(\theta;1)=\frac{1}{2}(\cos(\alpha\pi)-\cosh\theta)\,,
\eeq
and the fact that the functions $f(\theta,\alpha;n)$ all have leading behaviour $\Or((n-1)^0)$ as $n$ tends to 1.
Employing once more the improved asymptotics we have obtained the results of  Table~3 and Table~5. It is important to note, that it was crucial to have an explicit solution  for the coefficients to extract the $n\to 1$ limit, since the fit from different values of $1<n<2$ can give misleading coefficients in certain cases. 
\begin{table}[h!]
\[
\begin{array}{|c||c|}
\hline A_{11}(n) & 0.2036599645689198+\Or \left(n-1\right) \\
\hline B_{11}(n) & 0.0418206628975086+\Or \left(n-1\right)\\
\hline C_{11}(n) & \Or \left(n-1\right)\\
\hline A_{12}(n) & -0.4583212393562862(n-1)+\Or \left((n-1)^{2}\right)\\
\hline B_{12}(n)& -0.0581657847796917(n-1)+\Or \left((n-1)^{2}\right)\\
\hline C_{12}(n) & \Or \left((n-1)^{2}\right)
\\\hline \end{array}
\]
\caption{\label{tab:coeff_limit}$n\to1$ leading behaviour of the coefficients of (\ref{q1}) and (\ref{q2}). Note that the precise coefficient of $(n-1)$ in $C_{11}(n)$ is not given because such term will give no overall contribution to the von Neumann entropy. The same applies to $C_{12}(n)$.}
\end{table}

We point out, that to evaluate integrals of the two-particle form factors, like in Eq. \eqref{firstq}, in a numerically stable way, we need to regularize the integrals in \eqref{eq:fmin_n_alpha} on the line as was presented above for $\tilde{I}(\theta,\alpha;n)$, instead of using the formula where the $f_0(\theta,0;n)$ factor was pulled out of the expression. The reason is that $\tilde{I}(\theta,\alpha;n)$ diverges around $\theta=0$, which is compensated by the $f_0(\theta,0;n)$ term leading to a finite result, however this makes the numerical evaluation unstable around $\theta=0$. 

For the spin field, the structure of the two-particle form factors is very similar to that of the twist field form factors but slightly simpler. The formulae for $F_{a_1a_2}^{\sigma}(\theta)$ were all given in \cite{DM, DS}. Here we will only recall
\beqa 
F_{11}^{\sigma}(\theta)=\frac{\bar{\sigma}\cos^2 \frac{\pi}{3} \cos^2 \frac{\pi}{5} \cos^2 \frac{\pi}{30}  Q^\sigma_{11}(\theta)}
{\prod\limits_{\alpha=\frac{2}{3},\frac{2}{5},\frac{1}{15}}B_\alpha(\theta;1)}\frac{f_{11}(\theta;1)}{f_{11}(i\pi;1)}\,, \label{genff222}
\eeqa
and
\beqa 
F_{12}^{\sigma}(\theta)=\frac{\bar{\sigma} \cos^2 \frac{3\pi}{10} \cos^2 \frac{2\pi}{5} \cos^2 \frac{7\pi}{30} \cos^2 \frac{2\pi}{15}Q^\sigma_{12}(\theta)}
{ \prod\limits_{\alpha=\frac{4}{5},\frac{3}{5},\frac{7}{15},\frac{4}{15}} B_\alpha(\theta;1)}\frac{f_{12}(\theta;1)}{f_{12}(i\pi;1)}\,. \label{genff333}
\eeqa 
In \cite{DS}  functions 
\beq
Q_{11}^\sigma(\theta)=c_{11}^1 \cosh \theta+ c_{11}^0\,, \qquad Q_{12}^\sigma(\theta)=c_{12}^2 \cosh^2 \theta+ c_{12}^1\cosh\theta+ c_{12}^0\,,
\eeq
were computed and the coefficients are
\beqa
&&c_{11}^1=-2.093102832, \quad c_{11}^0=-10.19307727, \quad c_{12}^2=-7.979022182\,,\nonumber\\
&& c_{12}^1=-71.79206351, \quad \mathrm{and}\quad c_{12}^0=-70.29218939\,.
\eeqa

 \subsection{Entanglement Dynamics after a Quench of the Longitudinal Field}
Following Eq.~(9) of the Letter and the formulae given in \cite{PQ1, ourIsing} we have that the change experienced by the one-point function of the branch point twist field after a quench of the longitudinal field takes the form 
\beqa
&&\bra \Omega| \TT_n (0,t)|\Omega\ket= {}_{\mathrm{post}}\bra \Omega| \TT_n (0,0)|\Omega\ket_{\mathrm{post}}+  \delta_\lambda n  \sum_{a=1}^8 \frac{2}{{m}_{0,a}^2} F_a^{\sigma} F_a^{\TT_n } \cos(m_a t)  \label{firstq}\\
&& + 
2 \delta_\lambda n  \sum_{a,b=1}^8\int_{-\infty}^\infty 
\frac{\mathrm{d}p_a \mathrm{d}p_b}{2\pi e_a e_b }\frac{\delta(p_a+p_b)}{e_a+e_b} \mathrm{Re}\left[[F_{ab}^{\sigma}(p_a,p_b)]^* F_{ab}^{\TT_n}(p_a,p_b) [e^{- i  (\tilde{e}_{a}+\tilde{e}_{b}) t}]\right]+\Or(\delta_{\lambda}^2)\,,\nonumber
\eeqa
where $\delta_\lambda$ is the small change of the QFT coupling constant $\lambda_2\propto h_x$. Here we also used the fact that the one particle form factors are real. We also remind that $\tilde{e}_a(p)=\sqrt{m_a^2+p^2}$, being $m_a$ the post-quench mass of the type-$a$ particle.

From this expression it is relatively straightforward to arrive at the formula for the change of the entanglement entropies given in the Letter, see Eq.~(9) there. We know from the definition that
\beq
S_n(t):=\frac{1}{1-n} \log\left(\epsilon^{\Delta_{\TT_n}} \bra \Omega| \TT_n (0,t)|\Omega\ket \right)\,,
\eeq
therefore
\beqa
S_n(t)-S_n(0)&:=&\frac{1}{1-n} \log\left( \frac{\bra \Omega| \TT_n (0,t)|\Omega\ket}{\bra \Omega| \TT_n (0,0)|\Omega\ket} \right) \nonumber\\ &=&\frac{1}{1-n} \log\left(1+ \frac{\bra \Omega| \TT_n (0,t)|\Omega\ket-\bra \Omega| \TT_n (0,0)|\Omega\ket}{\bra \Omega| \TT_n (0,0)|\Omega\ket} \right)\,,
\eeqa
and, at first order in perturbation theory
\beq
S_n(t)-S_n(0)\approx \frac{1}{1-n} \frac{\bra \Omega| \TT_n (0,t)|\Omega\ket-\bra \Omega| \TT_n (0,0)|\Omega\ket}{\bra \Omega| \TT_n (0,0)|\Omega\ket}\,.
\eeq
The quantity ${}_{\mathrm{post}}\bra \Omega| \TT_n (0,0)|\Omega\ket_{\mathrm{post}}$, appearing on the RHS of~\eqref{firstq} can be also expanded  in a power series of $\delta_{\lambda}$. From dimensional analysis and the mass-coupling relation (\ref{mc}) we have that 
\beq
\tau_n=\langle\Omega|\mathcal{T}_n(0,0)|\Omega\rangle=A_{\TT_n} \,\lambda_2^{\frac{\Delta_{\TT_n}}{2-\Delta_\sigma}}\,,
\eeq
where $A_{\TT_n}$ is a non-universal function of $n$. Similarly
\beqa
{}_{\mathrm{post}}\bra \Omega| \TT_n (0,0)|\Omega\ket_{\mathrm{post}}= A_{\TT_n} (\lambda_2+\delta_\lambda)^{\frac{\Delta_{\TT_n}}{2-\Delta_\sigma}}=
 \tau_n \, \left(1+\frac{\delta_\lambda}{\lambda_2} {\frac{\Delta_{\TT_n}}{2-\Delta_\sigma}} + O(\delta^2_\lambda)\right)\,.
\label{vacuum}
\eeqa
From (\ref{vacuum}) and (\ref{firstq}) it then follows at first order in $\delta_\lambda$
\beqa
\label{ee1}
&& S_n(t)-S_n(0)\approx \frac{1}{1-n} \frac{\delta_\lambda}{\lambda_2} {\frac{\Delta_{\TT_n}}{2-\Delta_\sigma}} + \frac{\delta_\lambda n}{1-n}  \sum_{a=1}^8 \frac{2}{{m}_{0,a}^2} F_a^{\sigma} \hat{F}_a^{\TT_n } \cos(m_a t) \\
& &+   
\frac{2 \delta_\lambda n}{1-n}  \sum_{a,b=1}^8\int_{-\infty}^\infty 
\frac{\mathrm{d}p_a \mathrm{d}p_b}{2\pi e_a e_b }\frac{\delta(p_a+p_b)}{e_a+e_b} \mathrm{Re}\left[[F_{ab}^{\sigma}(p_a,p_b)]^* \hat{F}_{ab}^{\TT_n}(p_a,p_b) [e^{- i  (\tilde{e}_{a}+\tilde{e}_b) t}]\right]+\cdots\nonumber\,,
\eeqa
where the form factors of the twist field are normalized by the pre-quench vacuum expectation value $\tau_n$.
 Analogously, we can normalize the $\sigma$ form factors by the pre-quench expectation value $\bar{\sigma}$ introduced earlier. Generally, this expectation value has the form $\bar{\sigma}=A_\sigma \, \lambda_2^{\frac{\Delta_\sigma}{2-\Delta_\sigma}}$, where $A_\sigma$ is  a known constant. Expressing the masses $m_{0,a}$ in the denominators of Eq.~\eqref{ee1} in terms of the coupling as well (i.e. recalling Eq.~\eqref{mc}) we end up with 
\beqa
&& S_n(t)-S_n(0)=\frac{1}{1-n} \frac{\delta_\lambda}{\lambda_2}\left[ {\frac{\Delta_{\TT_n}}{2-\Delta_\sigma}} +  n \,\mathcal{C}_\sigma  \sum_{a=1}^8 \frac{2}{{r}_a^2} \hat{F}_a^{\sigma} \hat{F}_a^{\TT_n } \cos(m_a t)\right. \\
& &\left. +    2 n \,\mathcal{C}_\sigma   \sum_{a,b=1}^8\int_{-\infty}^\infty 
\frac{\mathrm{d}p_a \mathrm{d}p_b}{2\pi e_a e_b }\frac{\delta(\hat{p}_a+\hat{p}_b)}{\hat{e}_a+\hat{e}_b} \mathrm{Re}\left[[\hat{F}_{ab}^{\sigma}(p_a,p_b)]^* \hat{F}_{ab}^{\TT_n}(p_a,p_b) e^{- i  (\tilde{e}_{a}+\tilde{e}_b) t}\right]+\dots \right]+\Or(\delta_{\lambda}^2)\,,\nonumber
\eeqa
 where 
 \beq
 \mathcal{C}_\sigma=\frac{A_\sigma}{\kappa^2}\,,
 \eeq
 is constant featuring in Eq.~(9) of the Letter, $\kappa$ is given in Eq.~\eqref{mc}, $A_\sigma=-1.277(2)$ has been determined for instance in \cite{Fateev:1997yg}, $r_a$  are the normalized masses defined earlier, and $\hat{e}_a$ and $\hat{p}_a$  are the relativistic energies and momentums (see below Eq.~\eqref{formf}) divided  by the mass $m_{0,1}$. The ellipsis denotes higher particle number terms that are still first order in $\delta_\lambda $. 

In the $E_8$ minimal Toda field theory the (post-quench)  particle masses are such that $m_5>m_1+m_2>m_4>2m_1$. For this reason, oscillations coming from the two-particle form factor involving only particle types one and two, will have smaller frequency than those coming 
from the one-particle form factors of particle type five. Therefore, the six contributions to our expansion which involve the six smallest oscillation frequencies are precisely those coming from the form factors we have reviewed above. It is against these six contributions, that we have compared our numerical results in the Letter. More explicitly, expressing everything in terms of rapidities they are
\beqa
\label{osci}
&& S_n(t)-S_n(0)=\frac{1}{1-n} \frac{\delta_\lambda}{\lambda_2}\left[ {\frac{\Delta_{\TT_n}}{2-\Delta_\sigma}} +  n \,\mathcal{C}_\sigma  \sum_{a=1}^4 \frac{2}{{r}_a^2} \hat{F}_a^{\sigma} \hat{F}_a^{\TT_n } \cos(r_a  m_1 t)\right. \\ 
& &+    2 n \,\mathcal{C}_\sigma  \int_{-\infty}^\infty 
\frac{\mathrm{d}\theta }{2\pi }\frac{1}{2 \cosh^2\theta} \mathrm{Re}\left[[\hat{F}_{11}^{\sigma}(2 \theta)]^* \hat{F}_{11}^{\TT_n}(2 \theta) e^{- 2 i m_1 t \cosh\theta }\right] \nonumber  \\ 
& &+    2 n \,\mathcal{C}_\sigma  \int_{-\infty}^\infty 
\frac{\mathrm{d}\theta }{2\pi }\frac{1}{\cosh\theta(\cosh\theta+r_2 \cosh\tilde\theta)} \nonumber  \\
&&\left. \quad   \times   \mathrm{Re}\left[[\hat{F}_{12}^{\sigma}(\theta-\tilde\theta)]^* \hat{F}_{12}^{\TT_n}(\theta-\tilde\theta) e^{- i m_1 t (\cosh\theta+r_2 \cosh\tilde\theta) }\right]+\dots \right]+\Or(\delta_{\lambda}^2)\,, \nonumber
\eeqa
where 
\beq
\tilde{\theta}:=-\sinh^{-1}\left(\frac{\sinh\theta}{r_2}\right)\,.
\eeq
The limit $n\rightarrow 1$ needed to compute the von Neumann entropy can also be performed with the results given in the previous sections. For instance, for the one-particle form factor contributions, we just need to replace the form factors with the functions $g_i$ defined in (\ref{ga}).
This formula  will give oscillatory terms of frequencies $m_1, m_2, m_3, 2m_1,m_4$ and $m_1+m_2$ and can easily be evaluated numerically. Terms coming from the one-particle form factor contributions give undamped oscillations, whereas contributions from two-particle form factors, will produced damped oscillations, similar to those found in \cite{ourIsing} for a different quench.
For large $t$ it is possible to extract the leading oscillatory part of the two-particle form factor contributions by stationary phase analysis. These terms are suppressed as $t^{-3/2}$.

Finally, the variation of the expectation value of the spin field after the quench can also be obtained by the same techniques and the expression is almost identical to (\ref{firstq})
\beqa
\label{osci2}
&&\bra \Omega| \sigma (0,t)|\Omega\ket-\bar{\sigma} = \bar{\sigma} \frac{\delta_\lambda}{\lambda_2} \left[ {\frac{\Delta_{\sigma}}{2-\Delta_\sigma}}+   \mathcal{C}_\sigma \sum_{a=1}^8 \frac{2}{{r}_{a}^2} |\hat{F}_a^{\sigma}|^2 \cos(m_{a} t) \right. \\
& &+ 
\left. 2  \mathcal{C}_\sigma  \sum_{a,b=1}^8\int_{-\infty}^\infty 
\frac{\mathrm{d}p_a \mathrm{d}p_b}{2\pi e_a e_b } \frac{ \delta(\hat{p}_a+\hat{p}_b)}{\hat{e}_a+\hat{e}_b} |\hat{F}_{ab}^{\sigma}(p_a,p_b)|^2 \cos((\tilde{e}_{a}+\tilde{e}_b) t)+\dots\right]+\Or(\delta_{\lambda}^2)\,.\nonumber
\eeqa
In Fig.~\ref{fig:magn} we compare the numerical results for the one-point function of $\sigma(0,t)$ against the analytical formula (\ref{osci2}), incorporating the first four one-particle and the first two two-particle contributions.

\section{Transverse Field Quench: Mass Quench in Ising Field Theory}
\label{exp}

Considering the Ising field theory with $\lambda_2=0$, the model can be described by a free Majorana fermion with mass $m_0=\lambda_1$. In \cite{ourIsing} we presented a study of the evolution of entanglement after  a mass quench in this model. 
As explained in \cite{ourIsing}, the linked cluster expansion of the quench one-point function developed in \cite{SE} generalizes to the branch point twist field as
\beq
\frac{\bra\Omega|\TT_n(0,t)|\Omega\ket}{\bra\Omega|\Omega\ket}=\tilde{\tau}_n \sum_{k,l=0}^\infty D_{2k,2l}(t)\,,
\label{eq:1pt_D_expansion}
\eeq
where $|\Omega\ket$ is the pre-quench vacuum expressed in the post-quench particle basis as recalled in Eq.~(10) of the Letter, $\tilde{\tau}_n$ is the post-quench expectation value of the branch point twist field, $D_{2k,2l}$ is the combination of the expansion coefficients
\beq
\bra\Omega|\TT_n(0,t)|\Omega\ket=\tilde{\tau}_n \sum_{k,l=0}^\infty C_{2k,2l}(t)\,,
\eeq
and 
\beq
\bra\Omega|\Omega\ket=\sum_{q=0}^\infty Z_{2q}\,,
\eeq
in the form 
\beq
D_{2k,2l}(t)= \sum_{p=0}^{\min(k,l)}\tilde{Z}_{2p}C_{2(k-p),2(l-p)}(t)\,,
\label{eq:D_defintion}
\eeq
and $\tilde{Z}_{2p}$ is the inverse of $Z_{2q}$ defined as $\sum_{q=0}^{\infty}Z_{2q}\cdot \sum_{p=0}^{\infty}\tilde{Z}_{2p}=1$. In \cite{ourIsing} we showed that (\ref{eq:1pt_D_expansion}) was in fact the expansion of the exponential of a Laurent expansion in powers of $t$, with highest power 1. Our proof however was only carried out for terms in the expansion of order $K^2$ where $K(\theta)$ is a known function that enters the definition of $C_{i,j}(t)$. Our aim here is to provide a complete proof of exponentiation. Namely, the statement that (\ref{eq:1pt_D_expansion}) is an exponential is a general one and can be shown at all orders in $K(\theta)$. 
The precise definitions are
\begin{align}
&\tilde{\tau}_nC_{2k, 2l}(t)= \frac{1}{k! l!} \sum_{i_1,\ldots, i_{k}=1}^n  \sum_{j_1,\ldots, j_{l}=1}^n  \nonumber\\
& \times \left[\prod_{s=1}^{k }\int_{0}^\infty \frac{d\theta'_{s}}{2\pi}K(\theta'_{s})^* e^{2 i t E(\theta'_{s})}\right] \left[\prod_{r=1}^{l}\int_{0}^\infty \frac{d\theta_{r}}{2\pi}K(\theta_{r})e^{-2  i t E(\theta_r)}\right]\nonumber\\
 & \times  {}_{n;i_1i_1\ldots i_{k} i_{k}} \bra \theta'_1,-\theta'_1, \ldots, \theta'_{k},-\theta'_{k}|\TT_n(0,0)| -\theta_{l},\theta_{l},\dots, -\theta_{1}, \theta_1 \ket_{j_{l}j_{l}\ldots j_1 j_1;n}\,,
 \end{align}
and 
\beqa
&& Z_{2q}= \frac{1}{(q!)^2} \sum_{i_1,\ldots, i_{q}=1}^n  \sum_{j_1,\ldots, j_{q}=1}^n\left[\prod_{s=1}^{q }\int_{0}^\infty \frac{d\theta'_{s} d\theta_s}{(2\pi)^2}K(\theta'_{s})^* K(\theta_{s})\right]\nonumber\\
 && \times  {}_{i_1i_1\ldots i_{q} i_{q}} \bra \theta'_1,-\theta'_1, \ldots, \theta'_{q},-\theta'_{q}| -\theta_{q},\theta_{q},\dots, -\theta_{1}, \theta_1 \ket_{j_{q}j_{q}\ldots j_1 j_1} \quad \mathrm{for}\quad  q>0\,,
\eeqa
with $Z_0=1$. Combining the form of the expansion with the properties of the form factors such as the crossing relation and the Pfaffian nature of the multi-particle form factors (see \cite{ourIsing} for details),  a ``connected" expansion for the coefficients $C_{2k,2l}(t)$ and $Z_{2q}$ is suggested which reads
\beqa 
C_{2k,2l}(t)  &=&  \sum_{ \{n_{i,j}\}}^{k,l} \prod_{i,j=0}^\infty \frac{\left(C_{2i,2j}^{c}(t)\right)^{n_{i,j}}}{n_{i,j}!}\,, \\
Z_{2q}  &=& \sum_{ \{\tilde{n}_{j}\}}^{q} \prod_{j=0}^\infty \frac{\left(Z_{2j}^{c}\right)^{\tilde{n}_{j}}}{\tilde{n}_{j}!}\,, 
\eeqa
where the summations go for non-negative integer partitions satisfying the constraints $\sum_{i,j=0}^\infty i \, n_{i,j} =k$, $\sum_{i,j=0}^\infty j \,n_{i,j} =l$, and $\sum_{i=0}^\infty i\, \tilde{n}_i=q$.

The inverse coefficients $\tilde{Z}_{2k}$ also admit  a connected expansion 
\beq
\tilde{Z}_{2p}  = \sum_{ \{\tilde{m}_{i}\}}^{p} \prod_{i=0}^\infty \frac{\left(-Z_{2i}^{c}\right)^{\tilde{m}_{i}}}{\tilde{n}_{i}!}\,, 
\label{eq:Ztilde_connected_expansion}
\eeq
with the constraint $\sum_{j=0}^\infty j\, \tilde{m}_j=p$ , that we show by evaluating  the inverse relation 
\beq
\sum_{q=0}^{\infty}Z_{2q}\cdot \sum_{p=0}^{\infty}\tilde{Z}_{2p} = \sum_{q=0}^{\infty}\sum_{\left\{ \tilde{n}_{i}\right\} }^{q}\prod_{i=0}^{\infty}\frac{\left(Z_{2i}^{c}\right)^{\tilde{n}_{i}}}{\tilde{n}_{i}!} \cdot \sum_{p=0}^{\infty}\sum_{\left\{ \tilde{m}_{j}\right\} }^{p}\prod_{j=0}^{\infty}\frac{\left(-Z_{2j}^{c}\right)^{\tilde{m}_{j}}}{\tilde{m}_{j}!}\,.
\eeq
Let us reorganize the terms  and group them according to number of the function $K$, i.e. according to the value $\Delta=k+p$
\beq
\sum_{\Delta=0}^{\infty}\sum_{k=0}^{\Delta}\sum_{\left\{ \tilde{n}_{i}\right\} }^{k}\prod_{i=0}^{\infty}\frac{\left(Z_{2i}^{c}\right)^{\tilde{n}_{i}}}{\tilde{n}_{i}!} \sum_{\left\{ \tilde{m}_{j}\right\} }^{\Delta-k}\prod_{j=0}^{\infty}\frac{\left(-Z_{2j}^{c}\right)^{\tilde{m}_{j}}}{\tilde{m}_{j}!}\,.
\eeq
The sum of the integer partitions $\{\tilde{n}_i\}$ and $\{\tilde{m}_j\}$ can be seen as a new partition $\{\tilde{s}_i\}$ with constraint $\sum_{i=0}^\infty i\, \tilde{s}_i=\Delta$ that suggest further reorganization of the series to 
\beq
\sum_{\Delta=0}^{\infty}\sum_{\left\{ \tilde{s}_{i}\right\} }^{\Delta}\prod_{i=0}^{\infty}\sum_{t_{i}=0}^{\tilde{s_{i}}}\frac{\left(Z_{2i}^{c}\right)^{t_{i}}}{t_{i}!}\frac{\left(-Z_{2i}^{c}\right)^{\tilde{s}_{i}-t_{i}}}{(\tilde{s}_{i}-t_{i})!}\,,
\eeq
that, according to the binomial theorem, is nothing else but 
\beq
\sum_{\Delta=0}^{\infty}\sum_{\left\{ \tilde{s}_{i}\right\} }^{\Delta}\prod_{i=0}^{\infty}\frac{\left[Z_{2i}^{c}-Z_{2i}^{c}\right]^{\tilde{s}_{i}}}{\tilde{s}_{i}!} = 1\,,
\eeq
proving the connected expansion of the inverse coefficients. 

Examining the connected coefficients $C_{2k,2l}^{c}(t)$, we see that only diagonal terms, where $k=l$, are singular, since the successive application of the crossing relation leads to $C_{2k,2k}^{c}(t)\sim Z_{2k}^c$ in these cases. To have these singularities visible, we reorganize the expansion for $C_{2k,2l}(t)$ into product of diagonal and non-diagonal coefficients as
\beq
C_{2k,2l}=\sum_{p=0}^{\min\left(k,l\right)}\left[\sum_{\left\{ \tilde{m}_{i}\right\} }^{p}\prod_{i=0}^{\infty}\frac{\left(C_{2i,2i}^{c}\right)^{\tilde{m}_{i}}}{\tilde{m}_{i}!}\right]\left[\sum_{\tiny{\begin{array}{c}
\left\{ n_{ij}\right\} \\
i\neq j
\end{array}}}^{k-p,l-p}\prod_{i,j=0}^{\infty}\frac{\left(C_{2i,2j}^{c}\right)^{n_{ij}}}{n_{ij}!}\right]\,.
\eeq
Plugging this back to \eqref{eq:D_defintion} combining with \eqref{eq:Ztilde_connected_expansion} leads to
\beqa
D_{2k,2l}&=&\sum_{q=0}^{\min\left(k,l\right)}\sum_{p=0}^{\min\left(k-q,l-q\right)}\left[\sum_{\left\{ \tilde{n}_{i}\right\} }^{q}\prod_{i=0}^{\infty}\frac{\left(-Z_{2i}^{c}\right)^{\tilde{n}_{i}}}{\tilde{n}_{i}!}\right]\\&&\times\left[\sum_{\left\{ \tilde{m}_{i}\right\} }^{p}\prod_{i=0}^{\infty}\frac{\left(C_{2i,2i}^{c}\right)^{\tilde{m}_{i}}}{\tilde{m}_{i}!}\right]\left[\sum_{\tiny{\begin{array}{c}
\left\{ n_{ij}\right\} \\
i\neq j
\end{array}}}^{k-q-p,l-q-p}\prod_{i,j=0}^{\infty}\frac{\left(C_{2i,2j}^{c}\right)^{n_{ij}}}{n_{ij}!}\right]\,.
\eeqa
Introducing the variable $\Lambda=p+q$ we can rearrange the expression to 
\beqa
D_{2k,2l}&=&\sum_{\Lambda=0}^{\min\left(k,l\right)}\left\{ \sum_{r=0}^{\Lambda}\left[\sum_{\left\{ \tilde{n}_{i}\right\} }^{r}\prod_{i=0}^{\infty}\frac{\left(-Z_{2i}^{c}\right)^{\tilde{n}_{i}}}{\tilde{n}_{i}!}\right]\left[\sum_{\left\{ \tilde{m}_{i}\right\} }^{\Lambda-r}\prod_{i=0}^{\infty}\frac{\left(C_{2i,2i}^{c}\right)^{\tilde{m}_{i}}}{\tilde{m}_{i}!}\right]\right\} \\&&\times\left[\sum_{\tiny{\begin{array}{c}
\left\{ n_{ij}\right\} \\
i\neq j
\end{array}}}^{k-\Lambda,l-\Lambda}\prod_{i,j=0}^{\infty}\frac{\left(C_{2i,2j}^{c}\right)^{n_{ij}}}{n_{ij}!}\right]\,.
\eeqa
By a similar argument as above, we can argue that the terms inside the curly bracket evaluate to 
\beq
\sum_{\left\{ \tilde{s}_{i}\right\} }^{\Lambda}\prod_{i=0}^{\infty}\frac{\left(C_{2i,2i}^{c}-Z_{2i}^{c}\right)^{\tilde{s}_{i}}}{\tilde{s}_{i}!}\,.
\eeq
The combinations $D_{2i,2i}^{c}=C_{2i,2i}^{c}-Z_{2i}^{c}$ are regular, so are $D_{2i,2j}^{c}=C_{2i,2j}^{c}$  for $i\neq j$, hence the $D_{2k,2l}(t)$ also admits a connected expansion that is regular in the form 
\beq
D_{2k,2l}=\sum_{\left\{ n_{ij}\right\} }^{k,l}\prod_{i,j=0}^{\infty}\frac{\left(D_{2i,2j}^{c}\right)^{n_{ij}}}{n_{ij}!}\,.
\eeq
Combining this with \eqref{eq:1pt_D_expansion} leads to the proof of the exponential form of the one-point function
\beq
\frac{\bra\Omega|\TT_n(0,t)|\Omega\ket}{\bra\Omega|\Omega\ket}=\tilde{\tau}_n\exp\left[\sum_{k,l=0}^{\infty}D_{2k,2l}^{c}\right]\,.
\eeq
In the proof we did not use any special property of the branch point twist field form factors other than the crossing relation and the Pfaffian structure for the multi-particle form factors. These are true for other local operators as well, such as the spin field $\sigma$, hence we showed the full exponentiation of the one-point function of such operators too. 

\newpage

\section{Numerical Results}
\label{num}
The numerics in the present work were done using the infinite time evolving block decimation algorithm (iTEBD)~\cite{VidaliTEBD1,VidaliTEBD2}. Exploiting translational invariance, a general many-body state can be approximated by a matrix product state (written in the canonical form)
\begin{equation}
	\label{eq:MPS}
	|\Psi\ket = \sum_{\dots,s_j,s_{j+1},\dots}\dots \Lambda_{o}\Gamma_{o}^{s_j}\Lambda_{e}\Gamma_{e}^{s_{j+1}} \dots|\dots,s_j,s_{j+1},\dots \ket \,,
\end{equation}
where $\Gamma_{e/o}^s$ are $\chi\times\chi$ matrices associated with the even/odd lattice sites, $\Lambda_{e/o}$ are diagonal $\chi\times\chi$ matrices, with singular values $\lambda_i$ corresponding to the bipartition of the system along even/odd bonds.  The value of $\chi$ is the bond dimension. Expectation values of local operators can be calculated with standard tensor contraction procedures. The singular values on the bonds are the Schmidt coefficients corresponding to the bipartition, meaning that they are the eigenvalues of the reduced density matrix, therefore the entropies can be easily calculated.

\begin{figure}[htbp]
\centering
\includegraphics[width= 0.5 \columnwidth]{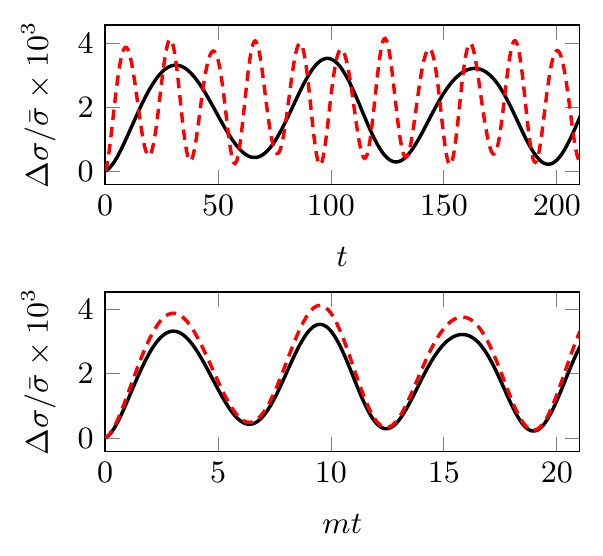}
\caption{The rescaling and mass coupling relation fit, for a quench with $\delta h_x/h_x = 0.05$ for the magnetization difference  $\Delta \sigma / \bar{\sigma} = \sigma(0,t)/\bar{\sigma}-1$. Solid curves correspond  to $h_x = 0.0005$ and dashed curves to $h_x = 0.005$. In the top plot the time is measured in ``proper" time, based on the energy scale defined by the lattice Hamiltonian Eq.(1) in the Letter, therefore the curve of larger $h_x$ (larger mass) has higher frequency oscillations. In the bottom we rescaled the time to be measured in units of $m^{-1}$, where $m\approx  5.42553(h_x+\delta h_x)^{8/15}$ to have the same frequency and allow for extrapolation. }
\label{fig:sig_resc}
\end{figure}

\begin{figure}[htbp]
\centering
\includegraphics[width=0.8 \columnwidth]{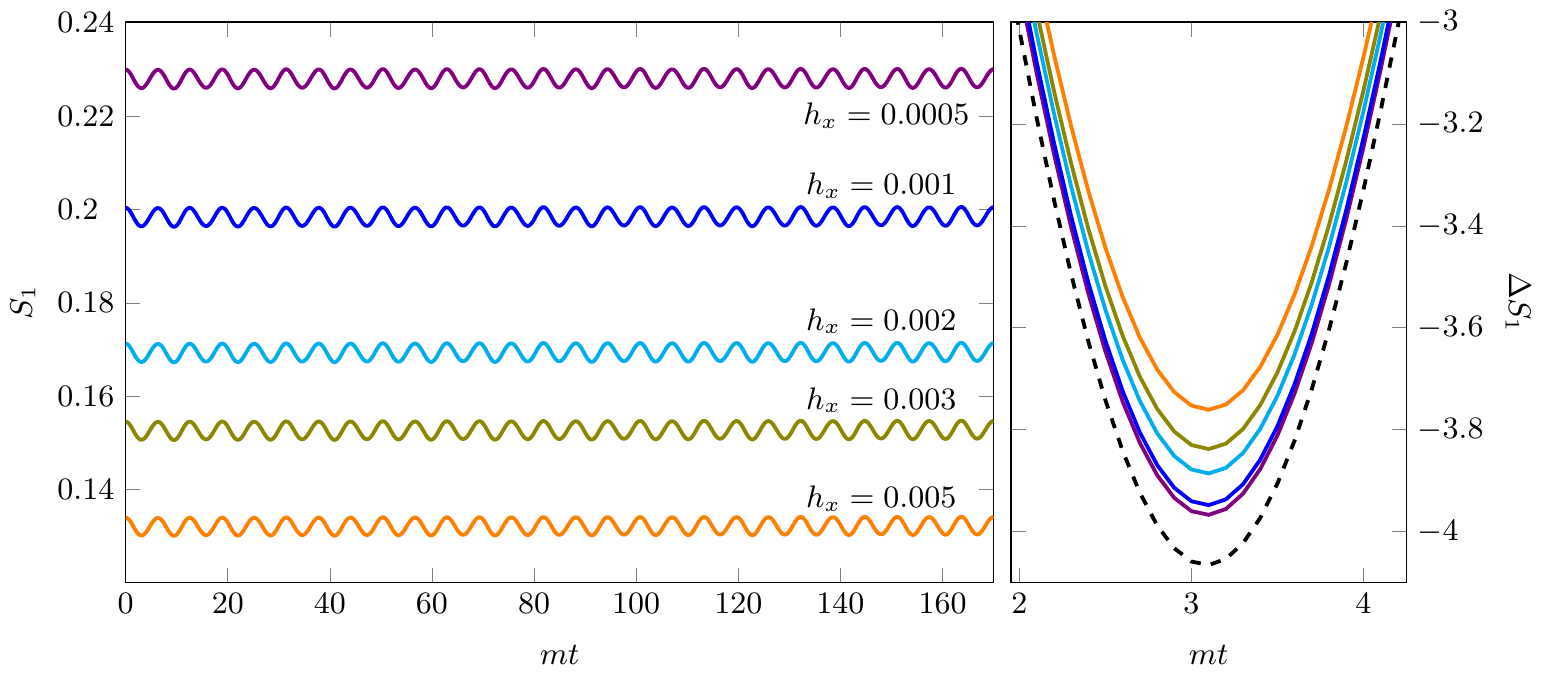}
\caption{Left panel: Results of the rescaling for the von Neumann entropy for $\delta h_x/h_x = 0.05$. In the time frame we have simulation data for extrapolation, there is no visible trace of entanglement growth. The time is rescaled to be measured in units of $m^{-1}$, where $m\approx  5.42553(h_x+\delta h_x)^{8/15}$. Right panel: Demonstration of the change of $\Delta S_1$ on a smaller time window, leading to convergence to the scaling limit.  The colour of the curves denote the same $h_x$ values as on the left panel. The dashed line is the extrapolated, scaling limit value of $\Delta S_1$, demonstrating the importance of calculating the scaling limit values to match the numerical data to the theoretical prediction.}
\label{fig:vn_nong}
\end{figure}

The simulation is based on the available code~\cite{Pollmann_notes}. 
In our adaptation, both for finding the initial state (pre-quench ground state) using imaginary time evolution, and for the real time evolution we used a fourth order Suzuki--Trotter decomposition~\cite{FOREST1990105} of the time evolution operator. For the imaginary time evolution, the time step was set to $\tau = 0.0005$ and we applied $N=200000$ Trotter steps, starting the iteration from the fully polarized state. For the post quench real time evolution the time step was set to $\delta t = 0.005$. We kept singular values $\lambda_i>10^{-12}$ up to a maximal bond dimension which was set to $\chi_{max}=300$. Due to the suppression of the entanglement growth shown in Fig.~\ref{fig:vn_nong} this was sufficient to carry out the simulations.  We used the same bond dimension for different couplings. As one gets closer to the critical point the von Neumann entropy of the initial state is expected to grow logarithmically with  the inverse mass. This was  perfectly captured by our  simulation that justifies the choice for the maximal bond dimension. We can also conclude that the suppression of the entanglement growth is not due to truncation effects  since states with higher entanglement are well approximated in our numerics.
We run simulations close to the critical point with couplings $h_z = 1$ (the critical value) and $h_x=0.0005, 0.001, 0.002, 0.003, 0.005$ for quenches with $\delta h_x/h_x = -0.04, 0.05$. The time dependent data has a leading frequency, corresponding to the mass of the lightest quasi-particle $m$. Due to dimensional analysis, rescaling the time with $m=\mathcal{B}_{lattice}(h_x+\delta h_x)^{8/15}$ one can plot the time signal in units of $m^{-1}$, i.e. all the time signals have ``leading" frequency $\tilde{\omega}=1$. For quenches with $\delta h_x/h_x = 0.05$ we obtained the fit $\mathcal{B}_{lattice}\approx 5.42553$, which was used for the  $\delta h_x/h_x = -0.04$ quenches as well for consistency, leading to the same period. This procedure is summarized in Fig.~\ref{fig:sig_resc} for the magnetization. In Fig.~\ref{fig:vn_nong} we show the result of the rescaling for the von Neumann entropy.

Once all the curves are scaled together, we approximate the data with interpolating curves. This allows us to extrapolate to the scaling limit $h_x \rightarrow 0$ up to $mt=170$. In the extrapolation procedure, first we determine the mass for each $h_x$, clearly having $m \rightarrow 0$ for decreasing $h_x$. Together with the time rescaling, this can be also interpreted as sending the lattice spacing ($a$) to zero, with fixed field theoretical mass $m = 1$. The mass and the lattice spacing always come in dimensionless combinations. We claim that one can extrapolate to the scaling limit using the scaling functions $S_n(a m) = \Delta_{\TT_n}/(1-n) \log (a m) + B (a m)^{1/n} + S_{n,\mathrm{scal.lim.}}$ for the entropies and  $(\Delta \sigma / \bar{\sigma})(a m) = \tilde{B} a m + (\Delta \sigma / \bar{\sigma})|_{\mathrm{scal.lim.}}$ for the magnetization for any value of $m t$ as in the transverse field case, see~\cite{CalCarPesUnusual,2010JSMTE..04..023C} and the Appendix of~\cite{ourIsing}. For the von Neumann entropy the prefactor of the logarithm term  is $-c/6$, and the fit of our data leads to $c\approx 0.49$ for all times less or equal than $mt=170$, and for both quenches studied.  This is  very close to the theoretical value $1/2$ which is a further indication that the maximal bond dimension  applied was sufficient to capture the  behaviour of the entanglement dynamics. For entropy differences between different times, the universal, time-independent logarithmic terms cancel, therefore we perform two parameter fits to extract the scaling limit values for $\Delta S_n$.  This process is displayed on the right hand side of Fig. \ref{fig:vn_nong} for the von Neumann entropy difference.

\begin{figure}[!hbp]
\centering
\includegraphics[width=0.5 \columnwidth]{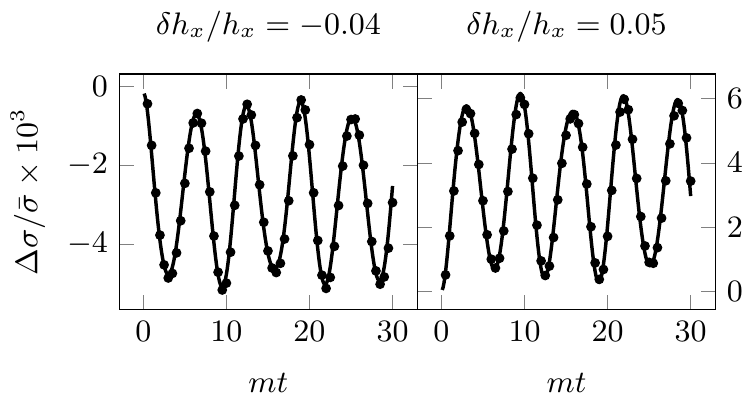}
\caption{The time evolution of the magnetizatoin for quenches with $\delta h_x/h_x = -0.04$ (left) and $\delta h_x/h_x = 0.05$ (right). The dots are the extrapolated iTEBD data. The lines are the theoretical predictions given in~\eqref{osci2}, incorporating the first four one-particle and the first two two-particle contributions.}
\label{fig:magn}
\end{figure}

\begin{figure}[!h!]
\centering
\includegraphics[width= 0.5 \columnwidth]{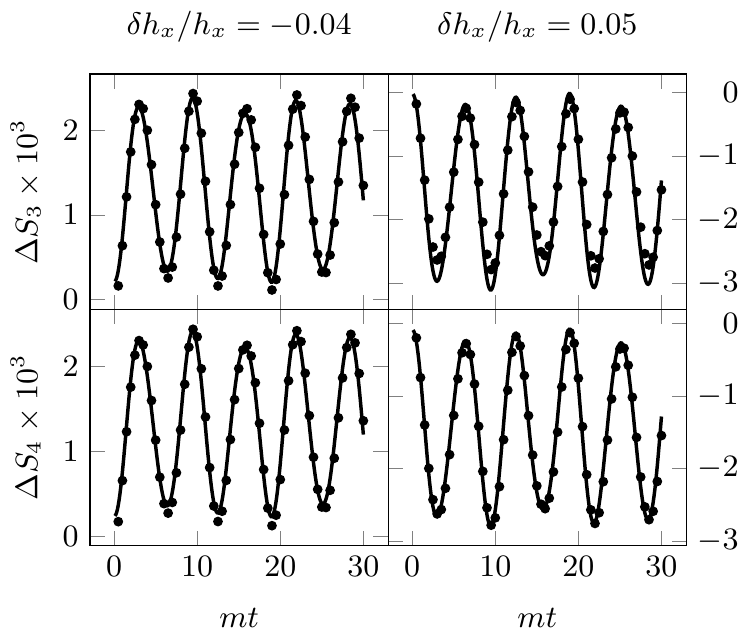}
\caption{The time evolution of the 3rd (top) and 4th (bottom) R\'enyi entropies for quenches with $\delta h_x/h_x = 0.05$ (left) and $\delta h_x/h_x = -0.04$ (right). The dots are the extrapolated iTEBD data. The lines are the theoretical predictions given in~\eqref{osci},  incorporating the first four one-particle and the first two two-particle contributions.}
\label{fig:r34}
\end{figure}

Additionally to the plots in the main text, here we plot our results for the initial time evolution of the magnetization in Fig.~\ref{fig:magn}, and for the third and fourth R\'enyi entropies in Fig.~\ref{fig:r34}. As in the main text for the von Neumann and the second R\'enyi entropies, we shifted the curves vertically  by empirical values to compensate for possible higher order corrections that are constant in time. We comment on these contributions on the next Section and list the values of the shifts applied in Table~\ref{tab:shifts}.

\begin{table}[h!]
\begin{center}

\end{center}
\end{table}

\begin{table}[h!]
\begin{center}
\begin{tabular}{ccccccc}
\hline
\multicolumn{1}{|c||}{$\delta h_x/h_x=\delta _\lambda/\lambda_2$}&\multicolumn{1}{|c|}{$\mathcal{D}_\sigma$}&\multicolumn{1}{|c|}{$\tilde{\mathcal{D}}_{\sigma}$}&\multicolumn{1}{||c|}{$\mathcal{D}_{1}$}&\multicolumn{1}{|c|}{$\tilde{\mathcal{D}}_{1}$}&\multicolumn{1}{||c|}{$\mathcal{D}_{2}$}&\multicolumn{1}{|c|}{$\tilde{\mathcal{D}}_{2}$}\\ \cline{1-7} \noalign{\vskip \doublerulesep \vskip-\arrayrulewidth} \cline{1-7}
\multicolumn{1}{|c||}{$-0.04$} & \multicolumn{1}{|c|}{$-0.00267$} & \multicolumn{1}{|c|}{$0.00005$}&\multicolumn{1}{||c|}{$0.00178$} &\multicolumn{1}{|c|}{$0.00015$}& \multicolumn{1}{||c|}{$0.00133$} & \multicolumn{1}{|c|}{$0.0001$}  \\ \hline
\multicolumn{1}{|c||}{$0.05$} & \multicolumn{1}{|c|}{$0.00333$}  &  \multicolumn{1}{|c|}{$-0.00025$} & \multicolumn{1}{||c|}{$-0.00222$} & \multicolumn{1}{|c|}{$0.00015$}  & \multicolumn{1}{||c|}{$-0.00167$} & \multicolumn{1}{|c|}{$0.00005$} \\ \cline{1-7} \noalign{\vskip \doublerulesep \vskip-\arrayrulewidth} \cline{1-5}
\multicolumn{1}{|c||}{$\delta h_x/h_x=\delta _\lambda/\lambda_2$ }&\multicolumn{1}{|c|}{$\mathcal{D}_{3}$}&\multicolumn{1}{|c|}{$\tilde{\mathcal{D}}_{3}$}&\multicolumn{1}{||c|}{$\mathcal{D}_{4}$}&\multicolumn{1}{|c|}{$\tilde{\mathcal{D}}_{4}$}&&\\ \cline{1-5} \noalign{\vskip \doublerulesep \vskip-\arrayrulewidth} \cline{1-5}
\multicolumn{1}{|c||}{$-0.04$} & \multicolumn{1}{|c|}{$0.00119$} & \multicolumn{1}{|c|}{$0.00015$} &\multicolumn{1}{||c|}{$0.00111$} &\multicolumn{1}{|c|}{$0.00025$} &&  \\  \cline{1-5}
\multicolumn{1}{|c||}{$0.05$} & \multicolumn{1}{|c|}{$-0.00148$}  &  \multicolumn{1}{|c|}{$0$} & \multicolumn{1}{||c|}{$-0.00139$} & \multicolumn{1}{|c|}{$-0.0001$} &&  \\   \cline{1-5}
\end{tabular}
\end{center}
\caption{\label{tab:shifts} The empirical vertical shifts applied to $\Delta \sigma/\bar{\sigma}$ and $\Delta S_n$ to compensate for possible higher order corrections that are constant in time are denoted by $\tilde{\mathcal{D}}_{\sigma}$ and $\tilde{\mathcal{D}}_{n}$ respectively. As comparison, we also list the value of the first order constant terms coming from the change of the expectation values in \eqref{osci2} and \eqref{ee1}, and we denoted them by $\mathcal{D}_\sigma = \delta_\lambda / \lambda_2\cdot\Delta_\sigma /(2-\Delta_\sigma)$ and  $\mathcal{D}_n = \delta_\lambda / \lambda_2\cdot\Delta_{\TT_n } /(2-\Delta_\sigma)/(1-n)$.
The applied shifts are one order of magnitude smaller than the first order result, in agreement with  the expectation.}
\end{table}

\section{Beyond First Order Perturbation Theory}
\label{bfo}

Finally we present some remarks on the suppression of linear growth and the constant in time shifts beyond first order perturbation theory. 

First of all we comment on the empirical vertical shifts applied to our results  summarized in Table \ref{tab:shifts}. The vertical offset is related to corrections to the VEV of local operators given in Eq.~\eqref{vacuum} up to the first order. However there are higher order corrections to the VEVs. These second order corrections were calculated for the transverse field Ising quench in~\cite{ourIsing}, and their magnitude is one order smaller to the first order values. Calculating the second order corrections to the offset for the longitudinal  quench  is an open problem beyond the scope of this paper,  hence they were calibrated by hand to match the numerical data to the theoretical prediction. The magnitudes of the these additional offsets compared to the first order results are one order smaller, as expected from the results of \cite{ourIsing}.

In~\cite{Overlap1}, a perturbative expansion for the pre-quench state $|\Omega\rangle$ was determined in the eigenstates of the post-quench Hamiltonian $H_{\text{post}}$ for the $E_8$ field theory up to $\mathcal{O}(\delta_{\lambda}^2)$; cf. Eq.~(4) of our Letter. Of particular importance to estimate the rate of entanglement growth are the overlaps of pairs with zero momentum~\cite{EEquench} 
 \begin{equation}
  K_{ab}(\theta):={}_{\text{post};a,b}\langle\theta,-\theta|\Omega\rangle\,;\quad a,b=1,\dots,8\,.
 \end{equation}

\begin{figure}[!htbp]
\centering
\includegraphics[width=  \columnwidth]{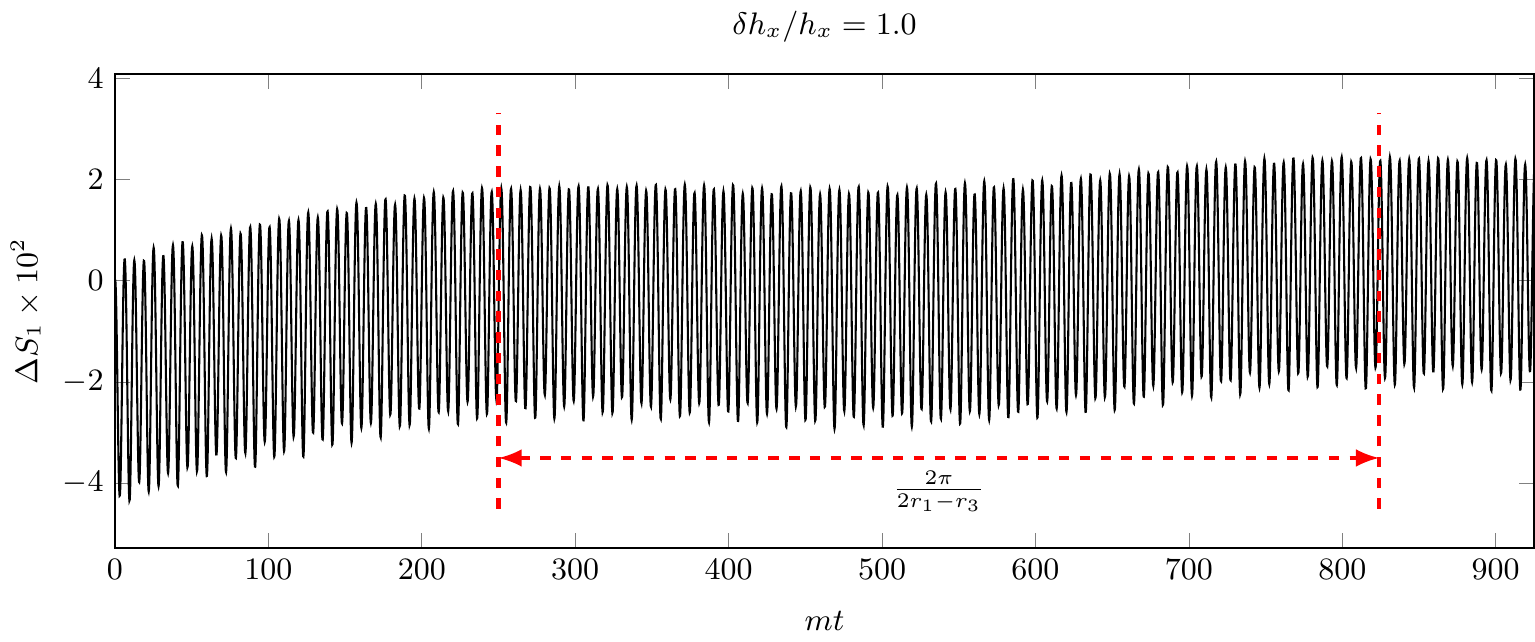}
\caption{Time evolution of entanglement entropy for a large  relative quench parameter $\delta h_x/h_x =1$  with  $h_x=0.005$. The data suggest a  slight drift in time of the entanglement entropy with oscillations present for large times as well. The low frequency modulation is related to the mass difference $2m_1-m_3$ that  is indicated by vertical dashed lines.}
\label{fig:big_quench}
\end{figure}

An interesting feature of these functions  can be inferred from plots presented in Section 4 of \cite{Overlap1}. One observes that the value of $|K_{11}(\theta_\star)|$ where $\theta_\star$ is the value of $\theta$ for which $|K_{11}(\theta)|$ is maximal, is of order $10^{-3}$ for $\delta_{\lambda}/\lambda_2=0.05$. In addition, $|K_{11}(\theta)|$ represents the largest particle overlap, so the maxima of other functions  
$K_{ab}(\theta)$ where $a,b$ are not both 1, are generally at least one order of magnitude smaller. If we compare these orders of magnitude with the same values for the function $|K(\theta)|$ involved in the study of the Ising field theory mass quench~\cite{SE}, we see that $|K(\theta_\star)| \approx 10^{-2}$ for the  same relative quench parameter, so roughly one order of magnitude larger. 
Since both the entanglement slope~\cite{ourIsing} and the higher order mass corrections~\cite{10.21468/SciPostPhys.5.3.027} are expected to be proportional to~  $|K_{ab}(\theta)|^2$, for small quenches in the $E_8$ field theory, they are small and not visible on the timescales accessible with iTEBD. In order to further support these claims we have run an additional simulation for a quench of $h_x = 0.005$, $\delta h_x = 0.005$ and very large times. The resulting time evolution is plotted in Fig. \ref{fig:big_quench}. The time rescaling was carried out using the same mass coupling relation as for small quenches. In this way the smallest frequency of the oscillations is close to one, within the resolution provided by the finite time window. This implies that even for a large quench, $\delta h_x/h_x=1$,  renormalization of the frequencies is absent. It is worth mentioning that the amplitudes of the  oscillations are not predictable from a first order perturbative calculation. However, the numerical results certainly suggest that the oscillations in the entanglement entropy are not suppressed by  higher order in perturbation theory. It is hard to assess whether the entropy will eventually grow linearly at large time from the numerical data, though failure to relax toward equlibrium  is clear. Finally we note that there is a very low frequency modulation of the signal, which turns out to be related to the  mass difference $2m_1-m_3$:  it originates from a second order contribution in perturbation theory, involving the operator matrix element between a one-particle state of particle $3$ and a two-particle state of particle $1$.

